\documentclass{tCON2e}

\usepackage{subfig}
\usepackage[usenames,dvipsnames]{pstricks}
\usepackage[permil]{overpic}
\usepackage[]{apacite}
\usepackage{placeins}

\usepackage[longnamesfirst,sort]{natbib}
\bibpunct[, ]{(}{)}{;}{a}{,}{,}
\renewcommand\bibfont{\fontsize{10}{12}\selectfont}

\theoremstyle{plain}

\theoremstyle{definition}

\theoremstyle{remark}

\begin{document}

\title{\Large{Grey-box State-space Identification \\ of Nonlinear Mechanical Vibrations}}

\author{
\name{J.P. No\"el\textsuperscript{a,b}$^{\ast}$\thanks{$^\ast$Corresponding author. Email: jp.noel@ulg.ac.be} and J. Schoukens\textsuperscript{b}}
\affil{\textsuperscript{a}Space Structures and Systems Laboratory \\ Aerospace and Mechanical Engineering Department \\ University of Li\`{e}ge, Li\`{e}ge, Belgium \\
\vspace{0.2cm}
\textsuperscript{b}ELEC Department \\ Vrije Universiteit Brussel, Brussels, Belgium}
}

\maketitle

\begin{abstract}
The present paper deals with the identification of nonlinear mechanical vibrations. A grey-box, or semi-physical, nonlinear state-space representation is introduced, expressing the nonlinear basis functions using a limited number of measured output variables. This representation assumes that the observed nonlinearities are localised in physical space, which is a generic case in mechanics. A two-step identification procedure is derived for the grey-box model parameters, integrating nonlinear subspace initialisation and weighted least-squares optimisation. The complete procedure is applied to an electrical circuit mimicking the behaviour of a single-input, single-output (SISO) nonlinear mechanical system and to a single-input, multiple-output (SIMO) geometrically nonlinear beam structure.
\end{abstract}


\rhead{\thepage}

\section{Introduction}

Operating in nonlinear regime has become a popular design solution for many systems and devices in order to meet escalating performance requirements. Among a variety of potential examples, the nonlinear implementation of signal processing operations is worthy of mention. In Ref.~[\cite{Daraio_AcousticSwitch}], tunable rectification was achieved using a granular crystal with bifurcating dynamics leading to quasiperiodic and chaotic states. By coupling nonlinear modes through internal resonances, Ref.~[\cite{Antonio_Stabilisation}] proposed a frequency stabilisation mechanism for micromechanical resonators. In Ref.~[\cite{Strachan_Cascade}], a chain of nonlinear resonators involving a cascade of parametric resonances was also used to passively divide frequencies. In the field of mechanical vibrations, intentionally utilising nonlinearity has similarly attracted a great deal of attention over the past few years, in particular for devising vibration absorbers~[\cite{Vakakis_NES,Habib_NLTVA}] and harvesters~[\cite{Karami_Pacemakers}].\\

This emergence of ever-more complicated nonlinear designs calls for the development of a new generation of data-driven modelling tools capable of accounting for nonlinear phenomena. The most recent and popular contributions pursuing this goal in mechanical vibration engineering were surveyed in Ref.~[\cite{Noel_Review}], and include, \textit{e.g.}, nonlinear modal analysis techniques~[\cite{Peeters_NPR_Part1,Renson_Continuation_NNM,Noel_NPS,Chen_VibroImpact}] and numerical model updating methods~[\cite{Sinou_HB,Cogan_HB}].\\

One of the major challenges in nonlinear data-driven modelling, or nonlinear system identification, is to address the constant compromise existing between the flexibility of the fitted model and its parsimony. Flexibility refers to the ability of the model to capture complex nonlinearities, while parsimony is its quality to possess a low number of parameters. In this paper, we concentrate on nonlinear state-space representations of the kind
\begin{equation}
\left\lbrace
\begin{array}{r c l}
    \mathbf{\dot{x}}(t) & = & \mathbf{A} \: \mathbf{x}(t) + \mathbf{B} \: \mathbf{u}(t) + \mathbf{E} \: \mathbf{g}(\mathbf{x},\mathbf{u},\mathbf{y}) \\
    \mathbf{y}(t) & = & \mathbf{C} \: \mathbf{x}(t) + \mathbf{D} \: \mathbf{u}(t) + \mathbf{F} \: \mathbf{h}(\mathbf{x},\mathbf{u},\mathbf{y}) ,
\end{array} \right.
\label{Eq:BlackBoxStateSpace}
\end{equation}
which can be classified as very flexible but little parsimonious, two features typically shared by black-box models. In Eqs.~(\ref{Eq:BlackBoxStateSpace}), $\mathbf{A} \in \mathbb{R}^{\: n_{s} \times n_{s}}$, $\mathbf{B} \in \mathbb{R}^{\: n_{s} \times m}$, $\mathbf{C} \in \mathbb{R}^{\: l \times n_{s}}$ and $\mathbf{D} \in \mathbb{R}^{\: l \times m}$ are the linear state, input, output and direct feedthrough matrices, respectively; $\mathbf{x}(t) \in \mathbb{R}^{\: n_{s}}$ is the state vector; $\mathbf{y}(t) \in \mathbb{R}^{\: l}$ and $\mathbf{u}(t) \in \mathbb{R}^{\: m}$ are the output and input vectors, respectively. The linear-in-the-parameters expressions $\mathbf{E} \: \mathbf{g}(\mathbf{x},\mathbf{u},\mathbf{y}) \in \mathbb{R}^{\: n_{s}}$ and $\mathbf{F} \: \mathbf{h}(\mathbf{x},\mathbf{u},\mathbf{y}) \in \mathbb{R}^{\: l}$ are the nonlinear model terms coupling the state, input and output variables. The order of the model, \textit{i.e.} the dimension of the state space, is noted $n_{s}$.\\

In the analysis of mechanical vibrations, one very often distinguishes nonlinearities distributed throughout (some large region of) the entire structure from localised nonlinearities, which are physically confined to a small area. Localised nonlinearities are arguably the most common in mechanical engineering practice, as they typically arise in joints interfacing substructures. Meaningful examples of this reality can be found in aerospace applications. During the modal survey of the Cassini spacecraft~[\cite{Carney_IMAC1997}], nonlinearities resulting from the appearance of gaps in the truss supports of the Huygens probe were attested. Similarly, the analysis of in-orbit data of the International Space Station highlighted that the opening of a pin connection in the assembly of its solar arrays led to severe nonlinear manifestations~[\cite{Laible_IMAC2013}]. Nonlinearities were also reported during ground vibration testing of the Airbus A400M, and were attributed to the elastomeric mounts supporting the four turboprop engines of the aircraft~[\cite{Ahlquist_A400M_IMAC2010}].\\

In the present paper, it is shown that, in the case of mechanical systems where nonlinearities are localised in physical space, the black-box model structure in Eqs.~(\ref{Eq:BlackBoxStateSpace}) can be drastically simplified. More specifically, Section~\ref{Sec:Model} demonstrates that the nonlinear terms in Eqs.~(\ref{Eq:BlackBoxStateSpace}) can be constructed using a limited number of output measurements, and hence without involving the state and input vectors. This makes the resulting grey-box state-space model a parsimonious representation of nonlinear mechanical systems. A two-step identification procedure is derived for this model in Section~\ref{Sec:ID}, integrating nonlinear subspace initialisation and weighted least-squares optimisation. Finally, the complete procedure is applied in Sections~\ref{Sec:Silverbox} and~\ref{Sec:NLBeam} to an electrical circuit mimicking the behaviour of a single-input, single-output (SISO) nonlinear mechanical system and to a single-input, multiple-output (SIMO) geometrically nonlinear beam structure, respectively.

\FloatBarrier
\newpage
\section{Grey-box state-space modelling based on Newton's second law}\label{Sec:Model}

\begin{figure}[ht]
\begin{center}
\scalebox{1.0} 
{
  \begin{pspicture}(11,4)
  \rput[bl](2.2,1){\includegraphics[width=0.40\textwidth]{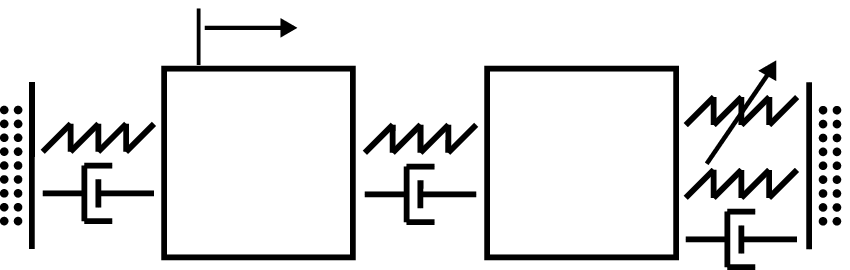}}
  \rput[bl](3.9,1.7){$m_1$}
  \rput[bl](6.5,1.7){$m_2$}
  \rput[bl](2.5,0.7){$k_{1}$,$c_{v1}$}
  \rput[bl](5.0,0.7){$k_{2}$,$c_{v2}$}
  \rput[bl](7.3,0.5){$c_{1}$,$k_{3}$,$c_{v3}$}
  \rput[bl](4,3.1){$u(t)$}
  \end{pspicture}
} 
\caption{An illustrative 2-degree-of-freedom mechanical system with one localised nonlinearity.}
\label{Fig:2DOF}
\end{center}
\end{figure}

Assuming localised nonlinearities, the vibrations of a $n_{p}$-degree-of-freedom mechanical system, \textit{i.e.} a system featuring $n_{p}$ linear resonances, obey Newton's second law written in the form
\begin{equation}
\mathbf{M} \: \mathbf{\ddot{y}}(t) + \mathbf{C}_{v} \: \mathbf{\dot{y}}(t) + \mathbf{K} \: \mathbf{y}(t) + \displaystyle \sum^{s}_{a=1} {c_{a} \: \mathbf{g}_{a}(\mathbf{y}_{nl}(t),\mathbf{\dot{y}}_{nl}(t))} = \mathbf{u}(t) ,
\label{Eq:Newton}
\end{equation}
where $\mathbf{M}$, $\mathbf{C}_{v}$, $\mathbf{K} \in \mathbb{R}^{\: n_{p} \times n_{p}}$ are the mass, linear viscous damping and linear stiffness matrices, respectively; $\mathbf{y}(t)$, $\mathbf{\dot{y}}(t)$, $\mathbf{\ddot{y}}(t)$ and $\mathbf{u}(t) \in \mathbb{R}^{\: n_{p}}$ are the displacement, velocity, acceleration and external force vectors, respectively; the nonlinear restoring force term is formed as the sum of $s$ basis function vectors $\mathbf{g}_{a}(t) \in \mathbb{R}^{\: n_{p}}$ associated with coefficients $c_{a}$. The subsets of displacements and velocities involved in the construction of the basis functions are denoted $\mathbf{y}_{nl}(t)$ and $\mathbf{\dot{y}}_{nl}(t)$, respectively. To fix ideas and illustrate Eq.~(\ref{Eq:Newton}), the 2-degree-of-freedom mechanical system shown in Fig.~\ref{Fig:2DOF} is considered. It comprises one cubic stiffness element, and obeys Newton's law as written term by term in Eq.~(\ref{Eq:2DOF}). Because the nonlinearity in the system is localised between mass 2 and a fixed base, vector $\mathbf{g}_{1} = \left(\begin{array}{c c} 0 & y^{3}_{2} \\ \end{array}	\right)^{T}$ in Eq.~(\ref{Eq:2DOF}) possesses a single nonzero element function of $y_{nl} = y_{2}$.\\

\vspace*{-0.5cm}

\vspace*{-1.0cm}
\begin{equation}
\begin{array}{r c l}
    \left(\begin{array}{c c} m_{1} & 0 \\
													   0 & m_{2} \\
					\end{array}	\right) \left(\begin{array}{c} \ddot{y}_{1} \\
																										 \ddot{y}_{2} \\
					\end{array}	\right) + \left(\begin{array}{c c} c_{v1}+c_{v2} & -c_{v2} \\
																												 -c_{v2} & c_{v2}+c_{v3} \\
					\end{array}	\right) \left(\begin{array}{c} \dot{y}_{1} \\
																										 \dot{y}_{2} \\
					\end{array}	\right) + & & \\
					 & & \\
					\left(\begin{array}{c c} k_{1}+k_{2} & -k_{2} \\
																												 -k_{2} & k_{2}+k_{3} \\
					\end{array}	\right) \left(\begin{array}{c} y_{1} \\
																										 y_{2} \\
					\end{array}	\right) + c_{1} \left(\begin{array}{c} 0 \\
																										 y^{3}_{2} \\
					\end{array}	\right) & = &
					\left(\begin{array}{c} u(t) \\
																 0 \\
					\end{array}	\right) \\
\end{array}
\label{Eq:2DOF}
\end{equation}

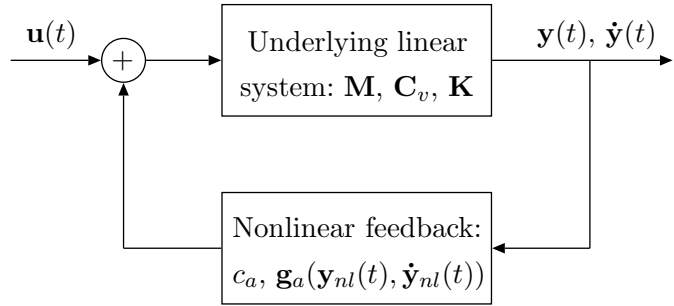
\begin{figure}[ht]
\begin{center}
    \begin{pspicture}(9,5)
    \psframe[linewidth=0.5pt,dimen=outer](2.8,0.5)(6.4,2) 
    \psframe[linewidth=0.5pt,dimen=outer](2.8,3.0)(6.4,4.5) 
		\psline[arrows=->,linewidth=0.5pt,arrowinset=0,arrowlength=1.2,arrowsize=3pt 2](0,3.75)(1.2,3.75) 
		\psline[arrows=->,linewidth=0.5pt,arrowinset=0,arrowlength=1.2,arrowsize=3pt 2](1.8,3.75)(2.8,3.75) 
		\psline[arrows=->,linewidth=0.5pt,arrowinset=0,arrowlength=1.2,arrowsize=3pt 2](6.4,3.75)(8.8,3.75) 
		\psline[arrows=->,linewidth=0.5pt,arrowinset=0,arrowlength=1.2,arrowsize=3pt 2](7.7,1.25)(6.4,1.25) 
		\psline[arrows=->,linewidth=0.5pt,arrowinset=0,arrowlength=1.2,arrowsize=3pt 2](1.5,1.25)(1.5,3.45) 
		\psline[linewidth=0.5pt](7.7,3.75)(7.7,1.25) 
    \psline[linewidth=0.5pt](2.8,1.25)(1.5,1.25) 
    \rput[bl](3.15,3.8){Underlying linear}
    \rput[bl](3.05,3.2){system: $\mathbf{M}$, $\mathbf{C}_{v}$, $\mathbf{K}$}
    \rput[bl](2.95,1.4){Nonlinear feedback:}
    \rput[bl](2.95,0.7){$c_{a}$, $\mathbf{g}_{a}(\mathbf{y}_{nl}(t),\mathbf{\dot{y}}_{nl}(t))$}
		\rput[bl](0.2,3.9){$\mathbf{u}(t)$}
    \rput[bl](7.0,3.9){$\mathbf{y}(t)$, $\mathbf{\dot{y}}(t)$}
		\pscircle[linewidth=0.5pt,dimen=outer](1.5,3.75){0.3}
    \rput[bl](1.34,3.62){+}
    \end{pspicture}
\caption{Feedback interpretation of Newton's law in Eq.~(\ref{Eq:Newton}).}
\label{Fig:Feedback}
\end{center}
\end{figure}

\vspace{1cm}The dynamics governed by Eq.~(\ref{Eq:Newton}) is conveniently interpreted by moving the nonlinear restoring force term to the right-hand side, \textit{i.e.}
\begin{equation}
\mathbf{M} \: \mathbf{\ddot{y}}(t) + \mathbf{C}_{v} \: \mathbf{\dot{y}}(t) + \mathbf{K} \: \mathbf{y}(t) = \mathbf{u}(t)  - \displaystyle \sum^{s}_{a=1} {c_{a} \: \mathbf{g}_{a}(\mathbf{y}_{nl}(t),\mathbf{\dot{y}}_{nl}(t))} ,
\label{Eq:Newton2}
\end{equation}
which leads to the block-diagram representation in Fig.~\ref{Fig:Feedback}.\\

The feedback structure of this diagram suggests that localised nonlinearities in mechanical systems act as additional inputs applied to the underlying linear system. This, in turn, reveals that black-box nonlinear terms in a state-space model, such as $\mathbf{E} \; \mathbf{g}(\mathbf{x},\mathbf{u},\mathbf{y})$  and $\mathbf{F} \; \mathbf{h}(\mathbf{x},\mathbf{u},\mathbf{y})$ in Eqs.~(\ref{Eq:BlackBoxStateSpace}), are overly complex to address mechanical vibrations. A more parsimonious description of nonlinearities is achieved by translating Eq.~(\ref{Eq:Newton2}) in state space, which provides the grey-box model
\begin{equation}
\left\lbrace
\begin{array}{r c l}
    \mathbf{\dot{x}}(t) & = & \mathbf{A} \: \mathbf{x}(t) + \mathbf{B} \: \mathbf{u}(t) + \mathbf{E} \: \mathbf{g}(\mathbf{y}_{nl}(t),\mathbf{\dot{y}}_{nl}(t)) \\
    \mathbf{y}(t) & = & \mathbf{C} \: \mathbf{x}(t) + \mathbf{D} \: \mathbf{u}(t) + \mathbf{F} \: \mathbf{g}(\mathbf{y}_{nl}(t),\mathbf{\dot{y}}_{nl}(t)) , \\
\end{array} \right.
\label{Eq:GreyBoxStateSpace0}
\end{equation}
where $\mathbf{g}(t) \in \mathbb{R}^{\: s}$ is a vector concatenating the nonzero elements in the basis function vectors $\mathbf{g}_{a}(t)$, and $\mathbf{E} \in \mathbb{R}^{\: n_{s} \times s}$ and $\mathbf{F} \in \mathbb{R}^{\: n_{p} \times s}$ are the associated coefficient matrices.\\

For the sake of conciseness, one adopts the concatenated equations
\begin{equation}
\left\lbrace
\begin{array}{r c l}
    \mathbf{\dot{x}}(t) & = & \mathbf{A} \: \mathbf{x}(t) + \overline{\mathbf{B}} \: \overline{\mathbf{u}}(t) \\
    \mathbf{y}(t) & = & \mathbf{C} \: \mathbf{x}(t) + \overline{\mathbf{D}} \: \overline{\mathbf{u}}(t) , \\
\end{array} \right.
\label{Eq:GreyBoxStateSpace}
\end{equation}
where $\overline{\mathbf{B}} = \left[\mathbf{B} \ \ \mathbf{E}\right]$ and $\overline{\mathbf{D}} = \left[\mathbf{D} \ \ \mathbf{F}\right]$; the extended input vector $\overline{\mathbf{u}}(t)$ is similarly defined as $\left[\mathbf{u}(t)^{T} \ \ \mathbf{g}(t)^{T}\right]^{T}$, where $T$ is the transpose operation.

\FloatBarrier
\newpage
\section{Identification procedure}\label{Sec:ID}

The simplified structure of the grey-box state-space model proposed in Eqs.~(\ref{Eq:GreyBoxStateSpace}) lends itself to a two-step identification procedure, whereas four steps including two nonlinear optimisation searches are needed in classical black-box state-space identification~[\cite{Paduart_PNLSS}]. First, initial estimates of the $\mathbf{A}$, $\overline{\mathbf{B}}$, $\mathbf{C}$ and $\overline{\mathbf{D}}$ matrices are obtained using nonlinear subspace identification. Second, the quality of the initial subspace parameter estimates is improved using nonlinear optimisation. The complete procedure is carried out in the frequency domain, opening the possibility to apply user-defined weighting functions in specific frequency intervals. We also opt for a discrete-time formulation to ensure a proper conditioning of the identification algorithm and to permit effective model-based time simulations.

\subsection{Initial state-space model obtained using nonlinear subspace identification}\label{Sec:Step1}

Eqs.~(\ref{Eq:GreyBoxStateSpace}) are first recast in the frequency domain as
\begin{equation}
\left\lbrace
\begin{array}{r c l}
    z_k \: \mathbf{X}(k) & = & \mathbf{A} \: \mathbf{X}(k) + \overline{\mathbf{B}} \: \overline{\mathbf{U}}(k) \\
    \mathbf{Y}(k) & = & \mathbf{C} \: \mathbf{X}(k) + \overline{\mathbf{D}} \: \overline{\mathbf{U}}(k) ,
\end{array} \right.
\label{Eq:FDStateSpace}
\end{equation}
where $k$ is the frequency line, $z_k$ the z-transform variable, and $\mathbf{Y}(k)$, $\mathbf{X}(k)$ and $\overline{\mathbf{U}}(k)$ the discrete Fourier transforms (DFTs) of $\mathbf{y}(t)$, $\mathbf{x}(t)$ and $\overline{\mathbf{u}}(t)$, respectively.\\

Initial estimates of the $\left(\mathbf{A},\overline{\mathbf{B}},\mathbf{C},\overline{\mathbf{D}} \right)$ matrices are calculated using the frequency-domain nonlinear subspace identification (FNSI) method proposed in Ref.~[\cite{Noel_FNSI}]. This method derives parameters non-iteratively from data by manipulating Eqs.~(\ref{Eq:FDStateSpace}) using geometric projections, and by considering the measured outputs as nonlinear regressors. The strength of using nonlinear subspace identification is that a fully nonlinear grey-box model is initially obtained, contrasting with the linearised initial model considered in the black-box state-space identification approach of Ref.~[\cite{Paduart_PNLSS}]. A complete description of the FNSI algorithm is provided in Appendix~1.

\subsection{Final state-space model obtained using nonlinear optimisation}\label{Sec:Step2}

The recourse to measured outputs as regressors in the FNSI method causes parameter estimates to suffer from a systematic error, which magnitude depends on the output signal-to-noise ratio (SNR). To reduce this error, the second step of the identification procedure consists in minimising a weighted least-squares cost function with respect to all parameters in $\left(\mathbf{A},\overline{\mathbf{B}},\mathbf{C},\overline{\mathbf{D}} \right)$.

\subsubsection{Definition of the cost function}

Introducing the vector of model parameters $\boldsymbol{\theta} \in \mathbb{R}^{\: n_{\theta}}$ as
\begin{equation}
\boldsymbol{\theta} = \left[vec\left(\mathbf{A}\right) \:;\: vec\left(\overline{\mathbf{B}}\right) \:;\: vec\left(\mathbf{C}\right) \:;\: vec\left(\overline{\mathbf{D}}\right) \right] ,
\label{Eq:theta}
\end{equation}
where the operation denoted $vec$ stacks the columns of a matrix on top of each other, the cost function to minimise writes
\begin{equation}
\mathbf{V}(\boldsymbol{\theta}) = \displaystyle \sum^{F}_{k=1} \boldsymbol{\epsilon}^{H}(k,\boldsymbol{\theta}) \: \mathbf{W}(k) \: \boldsymbol{\epsilon}(k,\boldsymbol{\theta}) ,
\label{Eq:V}
\end{equation}
where $F$ is the number of processed frequency lines, $H$ the Hermitian transpose, and $\mathbf{W}(k)$ a weighting function. The model error vector $\boldsymbol{\epsilon} \in \mathbb{R}^{\: l}$, where $l$ is in general the number of measured output variables, is defined as the complex-valued difference
\begin{equation}
\boldsymbol{\epsilon}(k,\boldsymbol{\theta}) = \mathbf{Y}_{m}(k,\boldsymbol{\theta}) - \mathbf{Y}(k) ,
\label{Eq:epsilon}
\end{equation}
where $\mathbf{Y}_{m}(k,\boldsymbol{\theta})$ and $\mathbf{Y}(k)$ are the DFTs of the modelled and measured outputs, respectively.\\

\subsubsection{Assumption on noise disturbances and weighting strategy}

An output error framework is adopted according to the two following assumptions on the frequency-domain input and output noise disturbances.\\

\textbf{Assumption 1}. The input spectrum is assumed to be noiseless, \textit{i.e.} observed without errors and independent of the output noise. In practice,
electromagnetic shakers used in mechanical applications typically yield SNRs of 60 to 80 $dB$, which is coherent with a noise-free assumption. If the input noise disturbances are otherwise too important, measurements can be averaged over multiple periods.\\

\textbf{Assumption 2}. The output disturbing noise term $\mathbf{N}_{Y}(k)$ is Gaussian distributed, has zero mean $\mathcal{E} \left (\mathbf{N}_{Y}(k) \right) = 0$, where $\mathcal{E}$ is the expectation operator, and has a covariance matrix with only nonzero diagonal elements equal to $\boldsymbol{\sigma}^{2}_{Y}(k) = \mathcal{E} \left ( \left| \mathbf{N}_{Y}(k) \right|^{2} \right)$, as described in Ref.~[\cite{Schoukens_Noise}].\\

In linear system identification, model errors are negligible relative to noise errors provided an adequate choice of the model order. In this case, the cost function $\mathbf{V}(\boldsymbol{\theta})$ in Eq.~(\ref{Eq:V}) is usually weighted by the inverse of the noise magnitude, \textit{i.e.} $\mathbf{W}(k) = \boldsymbol{\sigma}^{-1}_{Y}(k)$, leading to parameter estimates with maximum likelihood dispersion properties~[\cite{Schoukens_Book}]. In the presence of nonlinearities, model errors generally prevail over noise errors. Accordingly, we select herein a unitary weighting matrix $\mathbf{W}(k)$, assuming that unmodelled dynamics is uniformly distributed in the frequency domain.

\subsubsection{Analytical calculation of the Jacobian matrix}

In this work, the minimisation of the cost function $\mathbf{V}(\boldsymbol{\theta})$ in Eq.~(\ref{Eq:V}) is performed by means of a Levenberg-Marquardt optimisation algorithm, which combines the large convergence region of the gradient descent method with the fast convergence of the Gauss-Newton method~[\cite{Levenberg,Marquardt}]. This algorithm requires the calculation of the Jacobian matrix $\mathbf{J}(k,\boldsymbol{\theta})$ associated with the cost function or, equivalently, with the error function $\boldsymbol{\epsilon}(k,\boldsymbol{\theta})$, \textit{i.e.}
\begin{equation}
\mathbf{J}(k,\boldsymbol{\theta}) = \frac{\partial \boldsymbol{\epsilon}(k,\boldsymbol{\theta})}{\partial \boldsymbol{\theta}} = \frac{\partial \mathbf{Y}_{m}(k,\boldsymbol{\theta})}{\partial \boldsymbol{\theta}} .
\label{Eq:JWLS}
\end{equation}
Given the nonlinear relationship which exists between $\mathbf{Y}(k)$ and $\overline{\mathbf{U}}(k)$ in Eqs.~(\ref{Eq:FDStateSpace}), it may not be practical to compute the elements of $\mathbf{J}(k,\boldsymbol{\theta})$ in the frequency domain. An alternative approach consists in carrying out the computation of the Jacobian matrix in the time domain, and then in applying the DFT. Following this idea, the analytical derivation of all elements in $\mathbf{J}(k,\boldsymbol{\theta})$ is achieved in Appendix 2.

\FloatBarrier
\newpage
\section{Experimental demonstration on the Silverbox benchmark}~\label{Sec:Silverbox}

The identification procedure described in Section~\ref{Sec:ID} is demonstrated in the present section using experimental measurements acquired on the Silverbox benchmark, an electronic circuit mimicking the behaviour of a SISO mechanical system with a single resonance.\\

\begin{figure}[ht]
\begin{center}
\includegraphics[width=140mm]{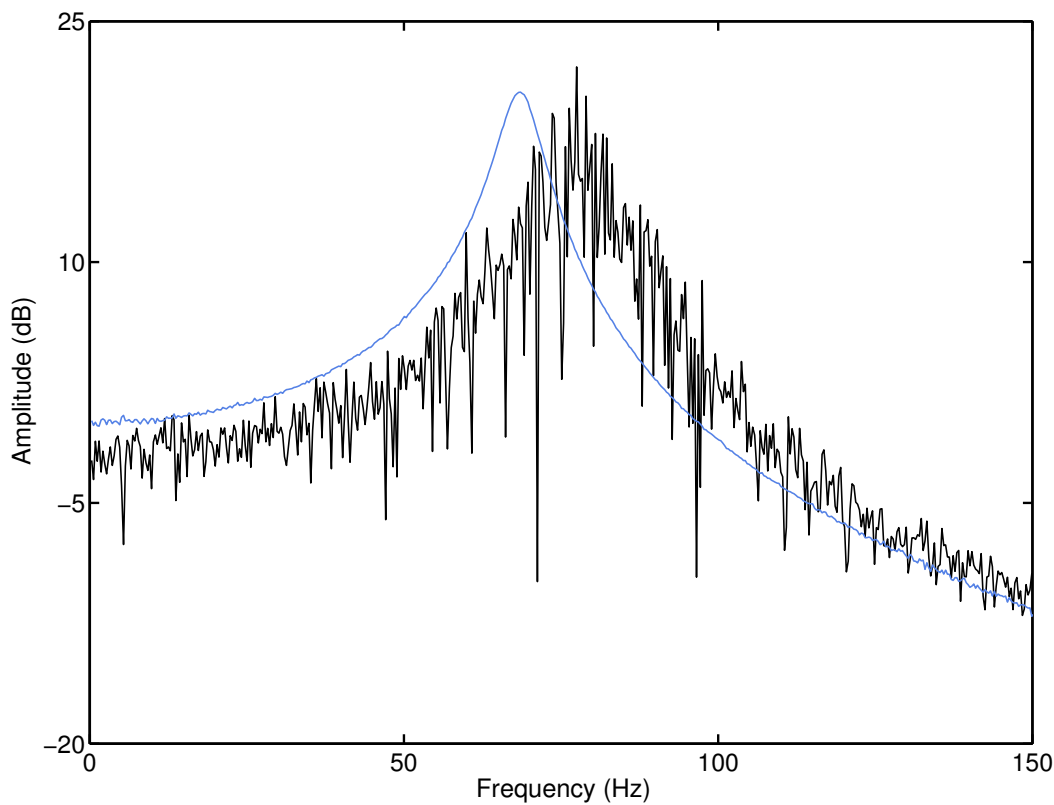} \\
\caption{Comparison of FRFs measured at 5 (in blue) and 100 (in black) $mV$ RMS.}
\label{Fig:Silverbox_FRF}
\end{center}
\end{figure}

The system was excited using random phase multisines~[\cite{Schoukens_Book}] with root-mean-squared (RMS) amplitudes of 5 and 100 $mV$. The input frequency spectrum was limited to 0 -- 300 $Hz$, excluding the DC component, and considering a sampling frequency of 2441 $Hz$. Experiments were conducted over 30 periods of 8192 samples each, removing the first 5 periods to achieve steady-state conditions. Table~\ref{Table:Silverbox} reports the underlying linear modal properties of the Silverbox estimated at 5 $mV$ RMS. Fig.~\ref{Fig:Silverbox_FRF} depicts frequency response functions (FRFs) measured at 5 and 100 $mV$ RMS. At high excitation level, a shift of the resonance frequency of about 9 $Hz$ together with severe nonlinearity-induced stochastic distortions are noticed.\\

\vspace*{0.2cm}\begin{table}[ht]
\begin{center}
\begin{tabular}{c c}
\hline
Natural frequency ($Hz$) & Damping ratio ($\%$) \\
68.58 &  4.68  \\
\hline
\end{tabular}
\caption{Natural frequency and damping ratio of the Silverbox benchmark estimated at 5 $mV$ RMS.} 
\label{Table:Silverbox}
\end{center}
\end{table}

\subsection{Grey- versus black-box identification}\label{Sec:Silverbox_GreyBlack}

The Silverbox was designed to exhibit the dynamics of a Duffing oscillator with cubic spring, though previous studies concluded that it also features an asymmetry in its nonlinear characteristic~[\cite{Schoukens_Silverbox,Noel_Silverbox}]. This is usually modelled using an additional quadratic stiffness term, prescribing the equation of motion
\begin{equation}
M \: \ddot{y}(t) + C_{v} \: \dot{y}(t) + K \: y(t) + c_{1} \: y^{2}(t) + c_{2} \: y^{3}(t) = u(t) .\\
\label{Eq:Duffing}
\end{equation}
A state-space model of the Silverbox system, in the grey-box form of Eqs.~(\ref{Eq:GreyBoxStateSpace0}), is constructed in discrete time at 100 $mV$ RMS. According to the physical description given in Eq.~(\ref{Eq:Duffing}), nonlinear terms in the state equation are chosen to be quadratic and cubic functions of the measured output displacement $y_{nl}(t)$. The resulting vector $\boldsymbol{\theta}$ in Eq.~(\ref{Eq:theta}) contains 13 parameters, given a model order equal to 2. In Ref.~[\cite{Paduart_PNLSS}], a black-box, second-order state-space model, in the discrete-time form equivalent to Eqs.~(\ref{Eq:BlackBoxStateSpace}), was identified considering a third-degree multivariate polynomial in the state equation with all cross products included, and linear terms only in the output equation. This led to a nonlinear model with 37 parameters.\\

\begin{figure}[p]
\begin{center}
\begin{tabular}{c}
\subfloat[]{\label{Fig:Silverbox_FDError}\includegraphics[width=140mm]{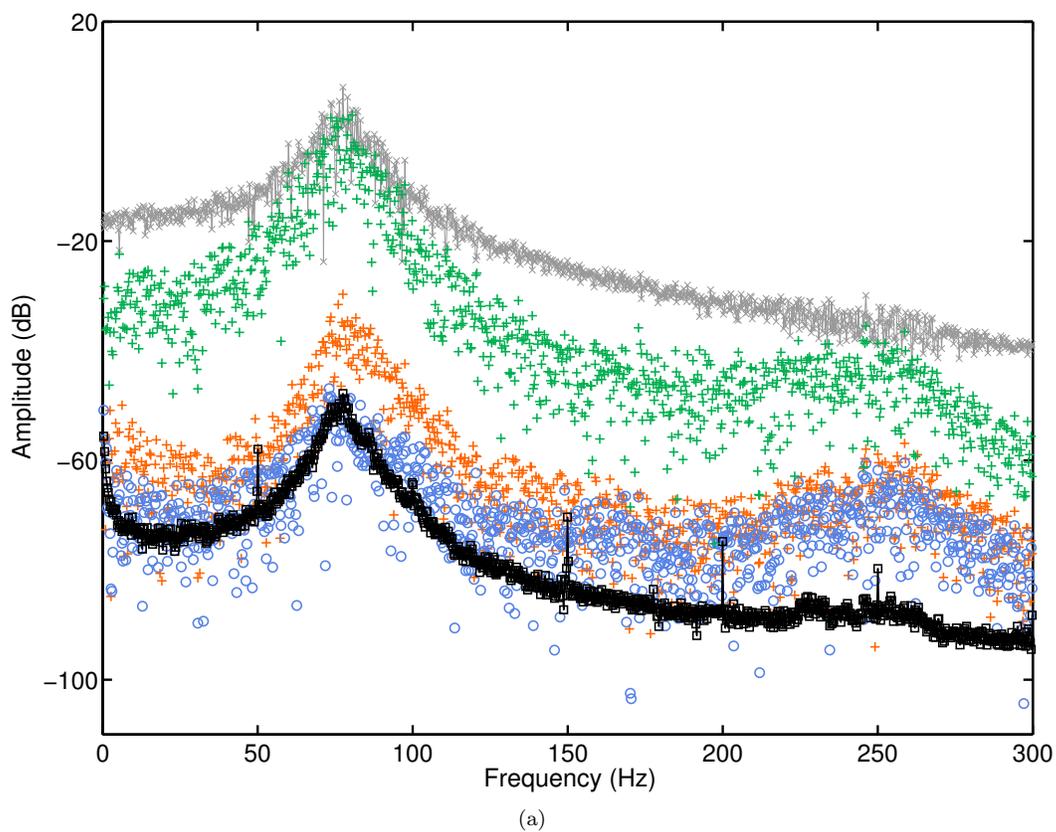}}\\
\subfloat[]{\label{Fig:Silverbox_TDError}\includegraphics[width=140mm]{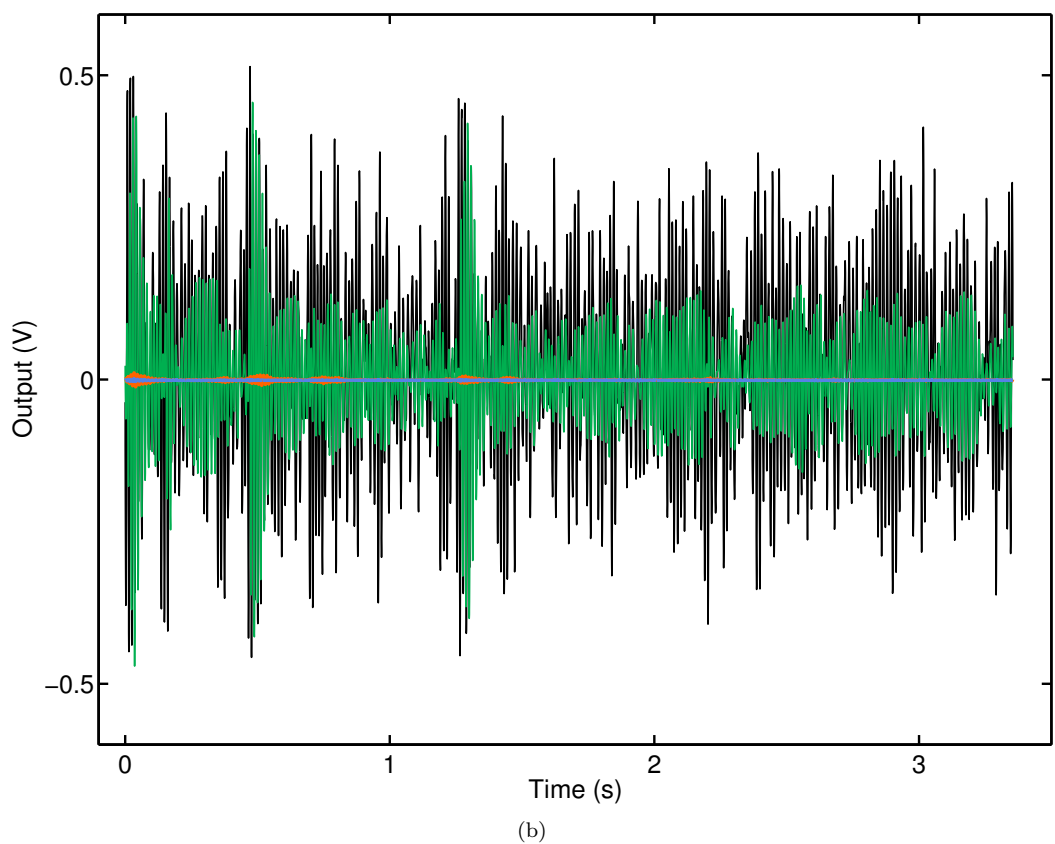}}\\
\end{tabular}
\caption{(a) Frequency-domain behaviour of the validation model error over the input band, featuring the output spectrum (in grey), linear state-space error level (in green), initial (in orange) and final (in blue) grey-box state-space error levels, and noise level (in black). (b) Time-domain validation plot, with the output time history (in black), linear state-space error (in green), and initial (in orange) and final (in blue) grey-box nonlinear state-space errors.}
\label{Fig:Silverbox_Error}
\end{center}
\end{figure}

The frequency-domain behaviour of the grey-box model error is studied in Fig.~\ref{Fig:Silverbox_Error}~(a), where the output spectrum in grey is compared with the initial and final state-space fitting error levels in orange and blue, respectively. The error of a linear state-space model is also plotted in green. Error levels were evaluated on a validation data set consisting of 1 period of 8192 samples measured at 100 $mV$ RMS. The noise level displayed in black was obtained by averaging estimation data over the 25 periods acquired in steady state. The initial model obtained using nonlinear subspace identification (in orange) features a low error level, generally lying 40 $dB$ below the output spectrum. This accuracy is explained by the high SNR of the output measurement (around 50 $dB$), which limits the systematic error discussed in Section~\ref{Sec:Step2}. The final model error (in blue) closely matches the noise level in the resonance region, translating a virtually perfect fitting in this interval. A larger error is noticed around 250 $Hz$, which might be attributed to an unmodelled resonance between the fifth harmonic of the noise signal and the third harmonic of the Silverbox response. The time-domain errors corresponding to Fig.~\ref{Fig:Silverbox_Error}~(a) are depicted in Fig.~\ref{Fig:Silverbox_Error}~(b). The RMS values of the validation output time history and of the linear model error are equal to 0.16 and 0.09 $V$, respectively. The initial and final state-space models obtained following the identification procedure of Section~\ref{Sec:ID} decrease this error down to 0.002 and 0.001 $V$, respectively.\\

\begin{figure}[p]
\begin{center}
\begin{tabular}{c}
\subfloat[]{\label{Fig:Silverbox_Initialisation}\includegraphics[width=140mm]{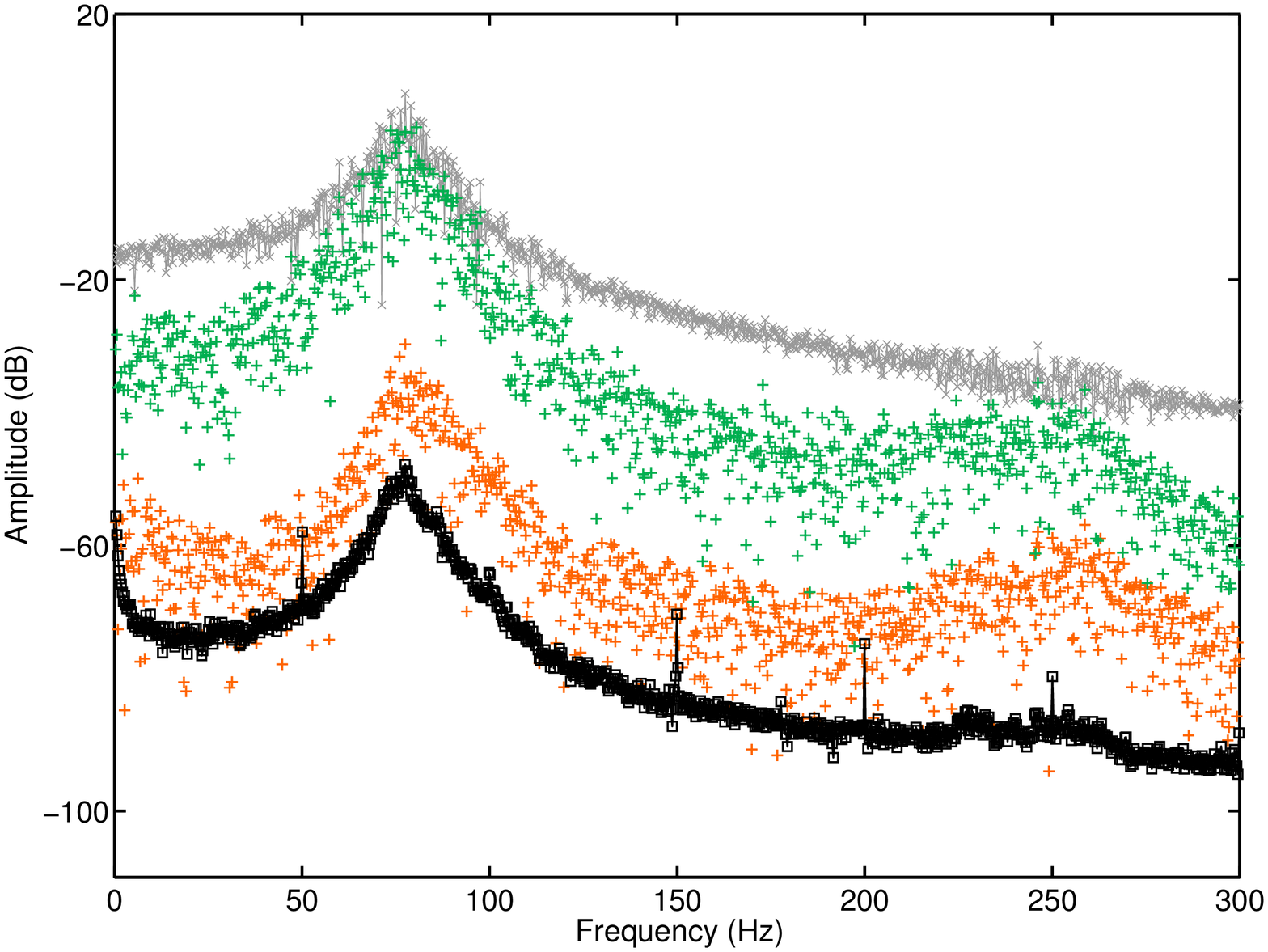}}\\
\subfloat[]{\label{Fig:Silverbox_GreyBlack}\includegraphics[width=140mm]{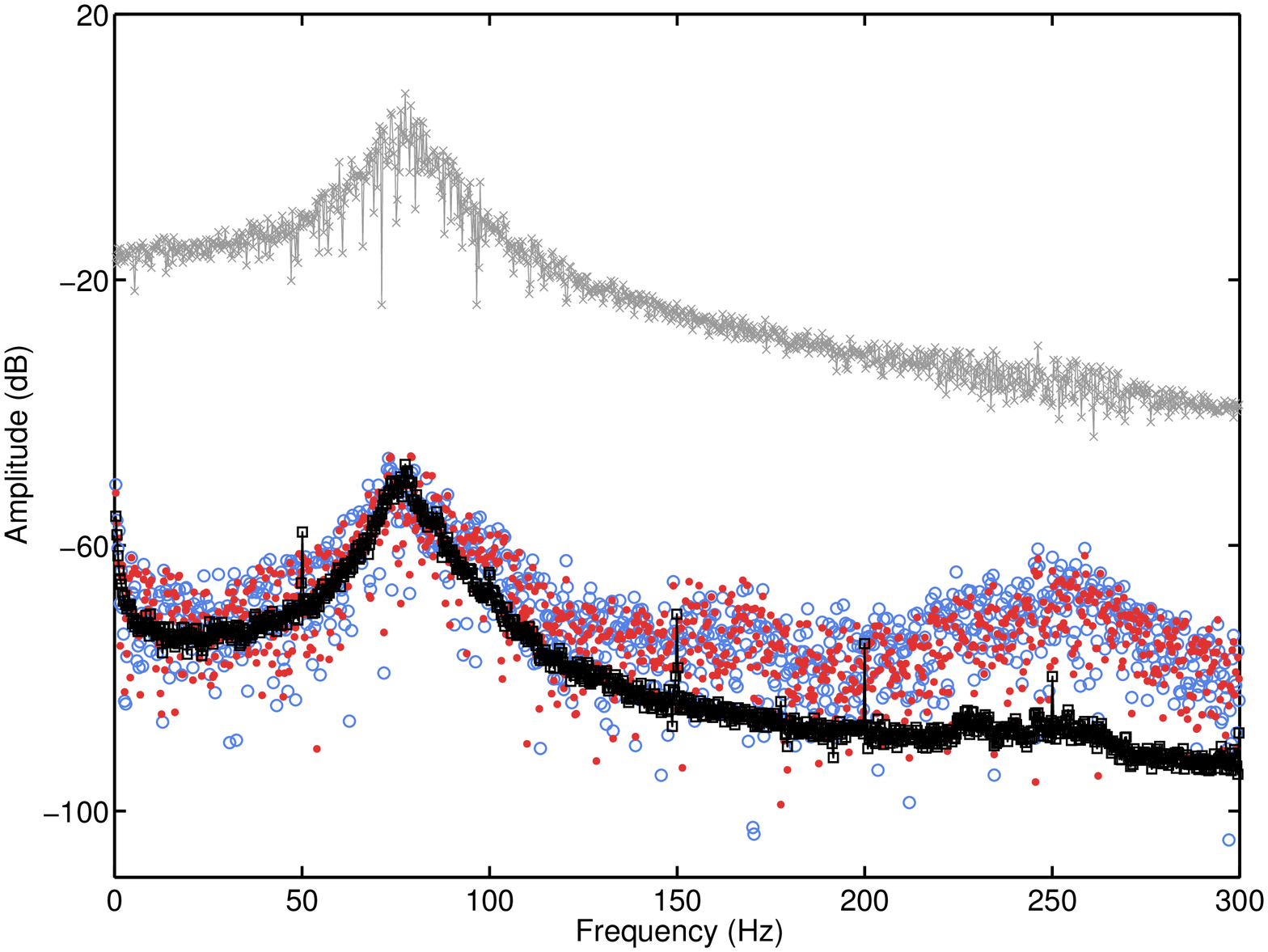}}\\
\end{tabular}
\caption{Frequency-domain comparison of the grey-box and black-box state-space modelling approaches. Validation output spectra and noise levels are plotted in grey and black, respectively. (a) Initial grey-box model obtained using nonlinear subspace identification (in orange) and initial linear black-box model (in green); (b) final grey-box (in blue) and black-box (in red) models.}
\label{Fig:Silverbox_Compare}
\end{center}
\end{figure}

\begin{figure}[ht]
\begin{center}
\includegraphics[width=140mm]{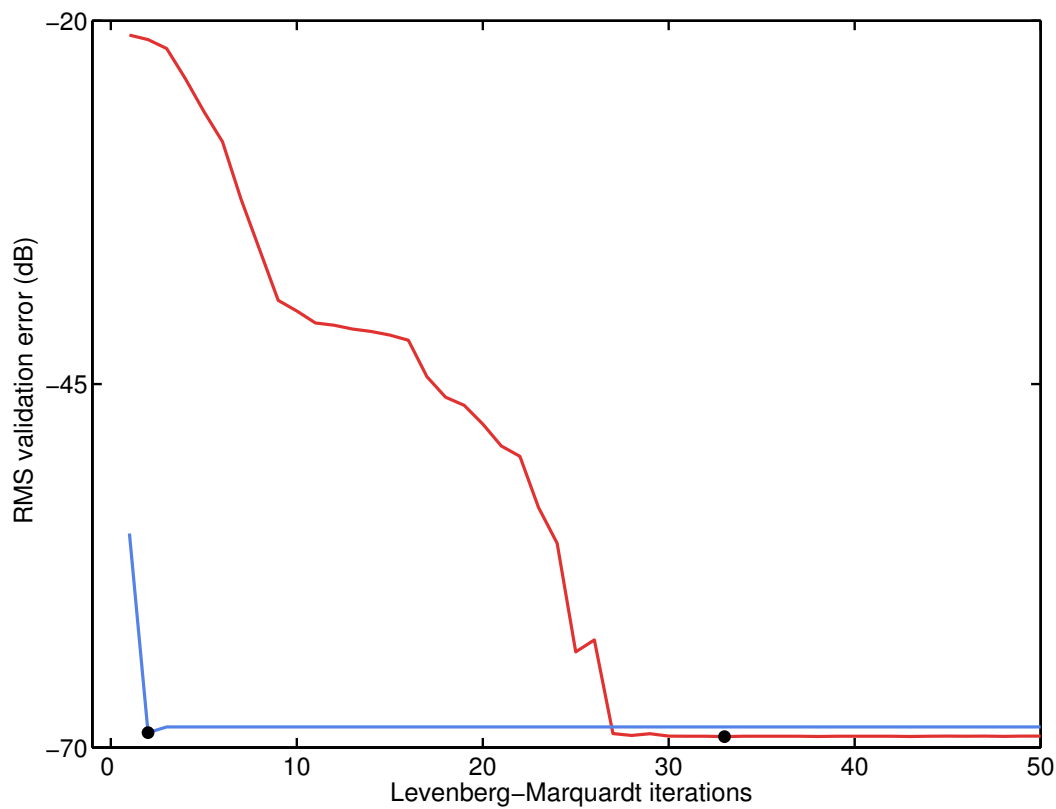} \\
\caption{Decrease of the RMS validation error over 50 Levenberg-Marquardt iterations for the grey-box (in blue) and black-box (in red) approaches. Optimal models are located using black dots.}
\label{Fig:Silverbox_Iterations}
\end{center}
\end{figure}

Fig.~\ref{Fig:Silverbox_Compare}~(a -- b) compares in the frequency domain the proposed grey-box approach with the black-box identification method of Ref.~[\cite{Paduart_PNLSS}]. In Fig.~\ref{Fig:Silverbox_Compare}~(a), initial model error spectra are superposed. The graph shows the clear advantage in starting from the nonlinear subspace model (in orange) described in Section~\ref{Sec:Step1}, as opposed to the linearised model (in green) adopted in black-box identification. Final models are assessed in Fig.~\ref{Fig:Silverbox_Compare}~(b). They are seen to lead to a similar validation error level, proving the accuracy of the introduced grey-box framework. The Levenberg-Marquardt iterations necessary to construct the final grey-box and black-box models are eventually analysed in Fig.~\ref{Fig:Silverbox_Iterations}. It is observed that an optimal grey-box model is reached in a single iteration, whereas 33 iterations are required by the black-box model. This significant difference is the result of the already low initial error of the nonlinear subspace model (-55 $dB$), in comparison with the more substantial error of the linearised model (-21 $dB$).

\newpage
\subsection{Nonlinear coefficients and discrete- to continuous-time conversion}\label{Sec:NLCoeff}

The coefficients of the physical system nonlinearities, \textit{i.e.} $c_{1}$ and $c_{2}$ in Eq.~(\ref{Eq:Duffing}), can be calculated from the final estimates of the grey-box state-space parameters. Following the procedure described in Ref.~[\cite{Marchesiello_TNSI}], the generated nonlinear coefficients are frequency-dependent and possess a spurious imaginary part. They are obtained as the ratios of elements of the transfer function matrix associated with the model in Eqs.~(\ref{Eq:FDStateSpace}). This matrix is formed in continuous time, assuming zero-order-hold equivalence between the discrete- and continuous-time domains.\\

\begin{figure}[p]
\begin{center}
\begin{tabular}{c c}
\subfloat[]{\label{Fig:Silverbox_QuadNL_Real}\includegraphics[height=56mm]{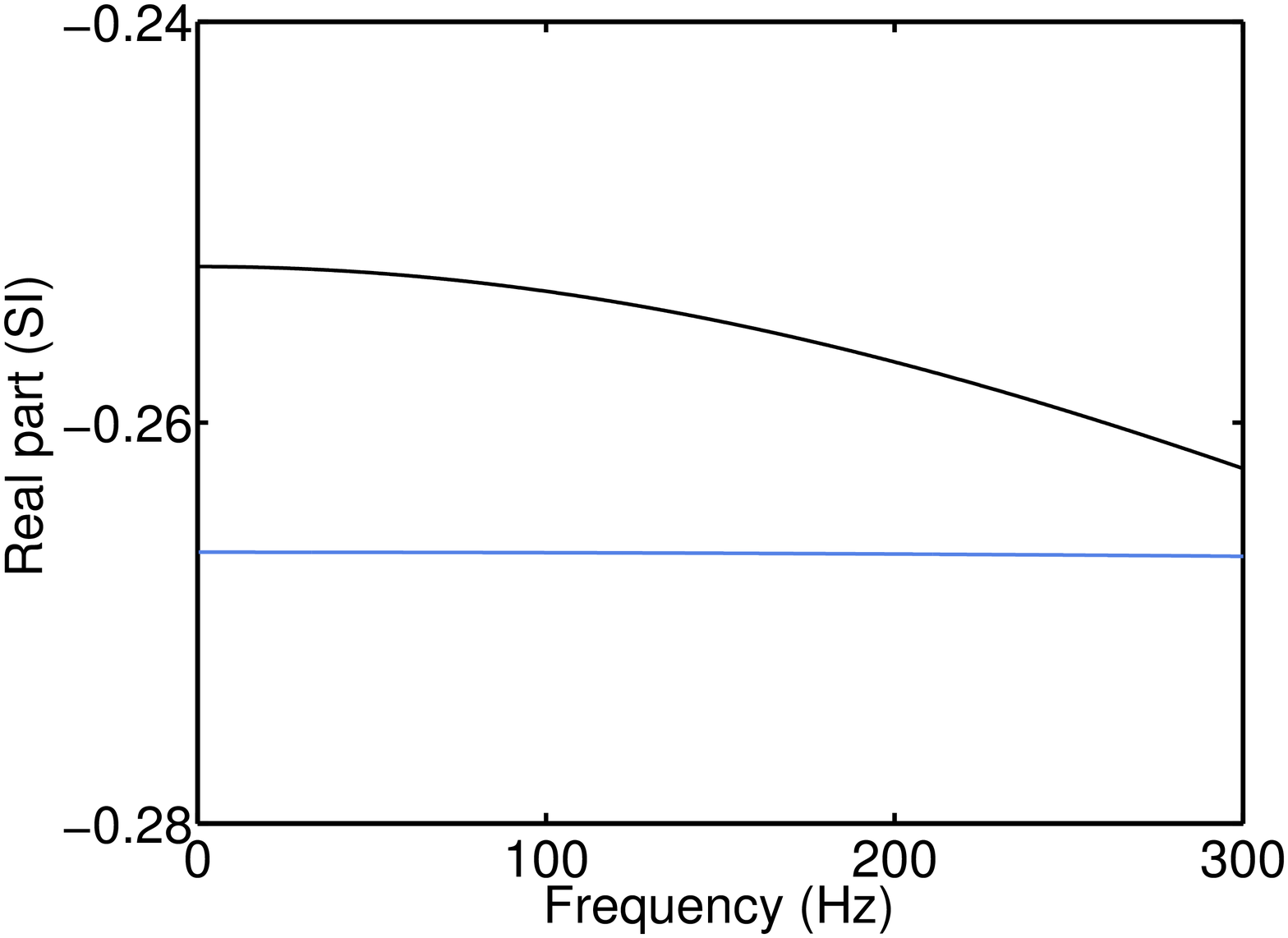}} &
\subfloat[]{\label{Fig:Silverbox_QuadNL_Imag}\includegraphics[height=58mm]{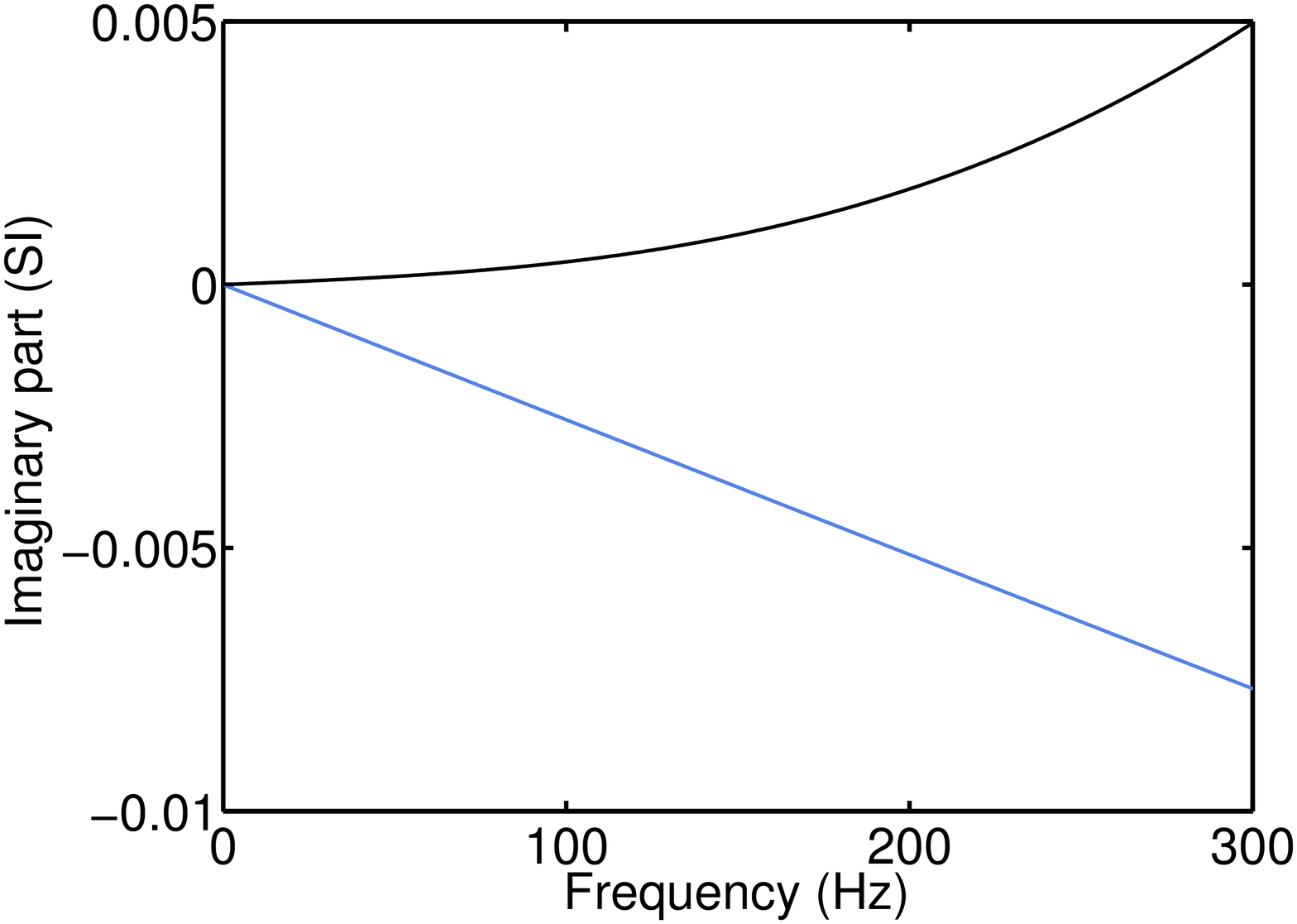}} \\
\subfloat[]{\label{Fig:Silverbox_CubicNL_Real}\includegraphics[height=56mm]{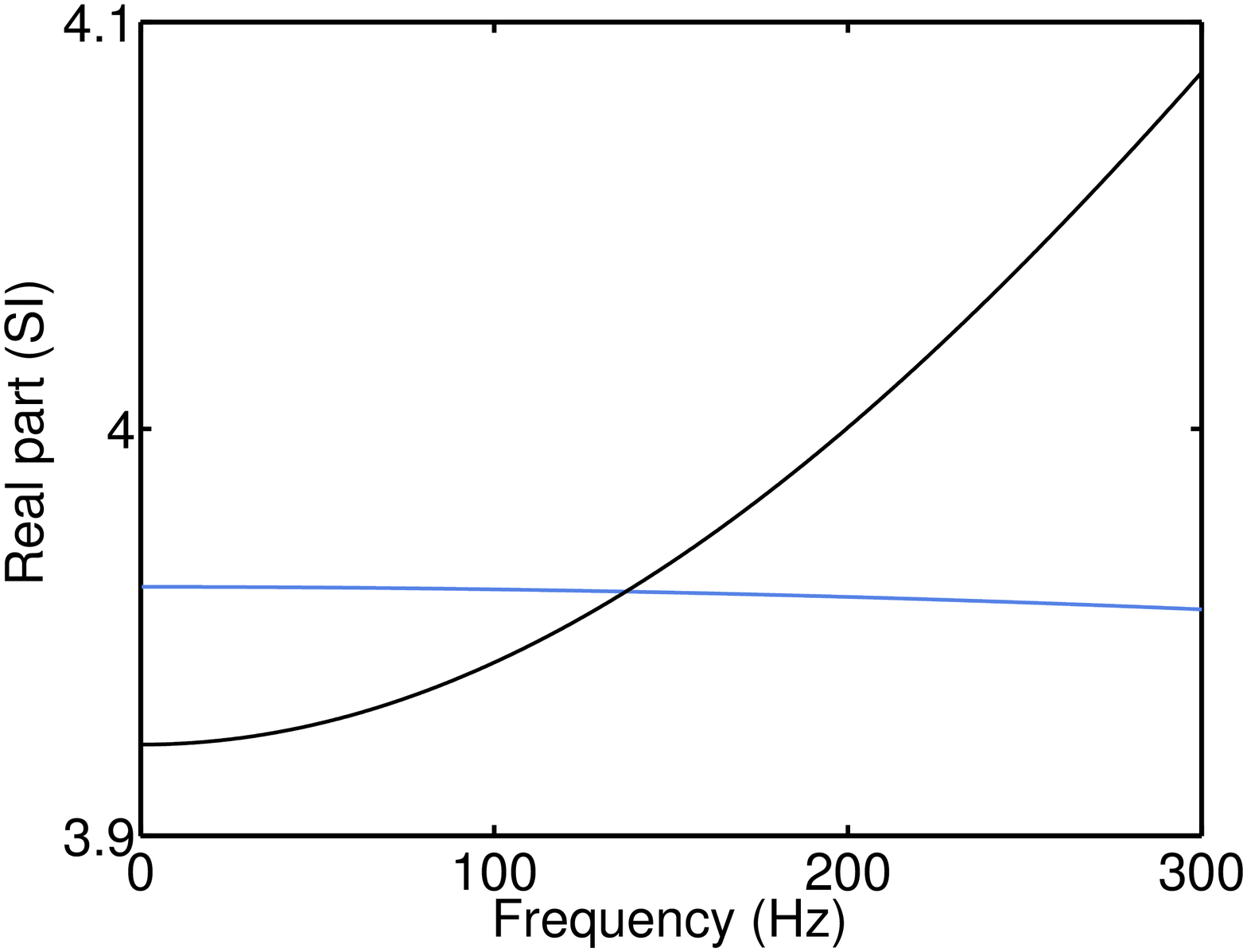}} &
\subfloat[]{\label{Fig:Silverbox_CubicNL_Imag}\includegraphics[height=55mm]{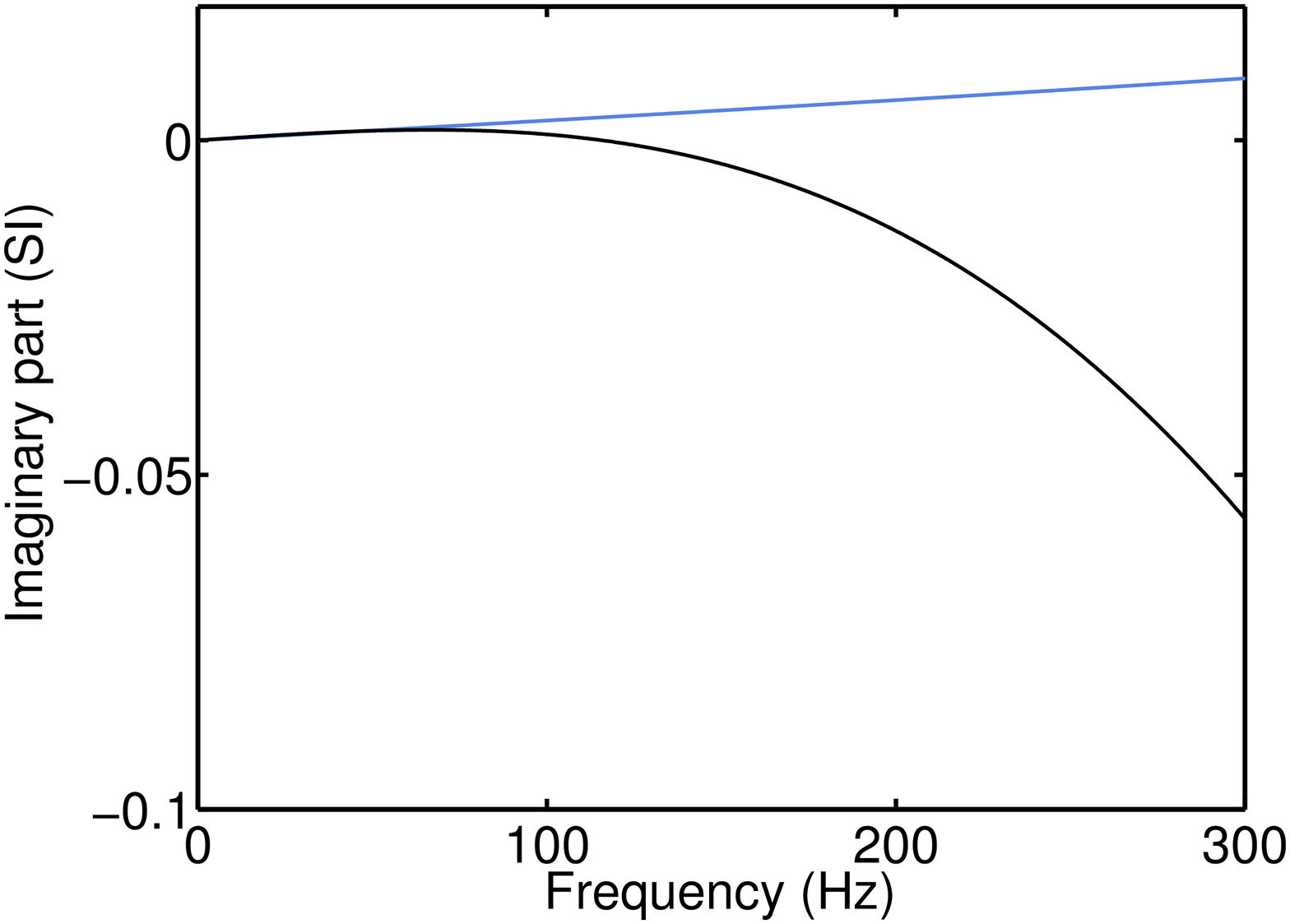}} \\
\end{tabular}
\caption{Complex-valued and frequency-dependent estimates of the nonlinear coefficients (a -- b) $c_{1}$ and (c -- d) $c_{2}$ for sampling frequencies of 2441 $Hz$ (in black) and 12205 $Hz$ (in blue).}
\label{Fig:Silverbox_NLCoeff}
\end{center}
\end{figure}

\begin{figure}[ht]
\begin{center}
\includegraphics[width=140mm,tics=50,trim={3.2cm 0 0 0},clip=true]{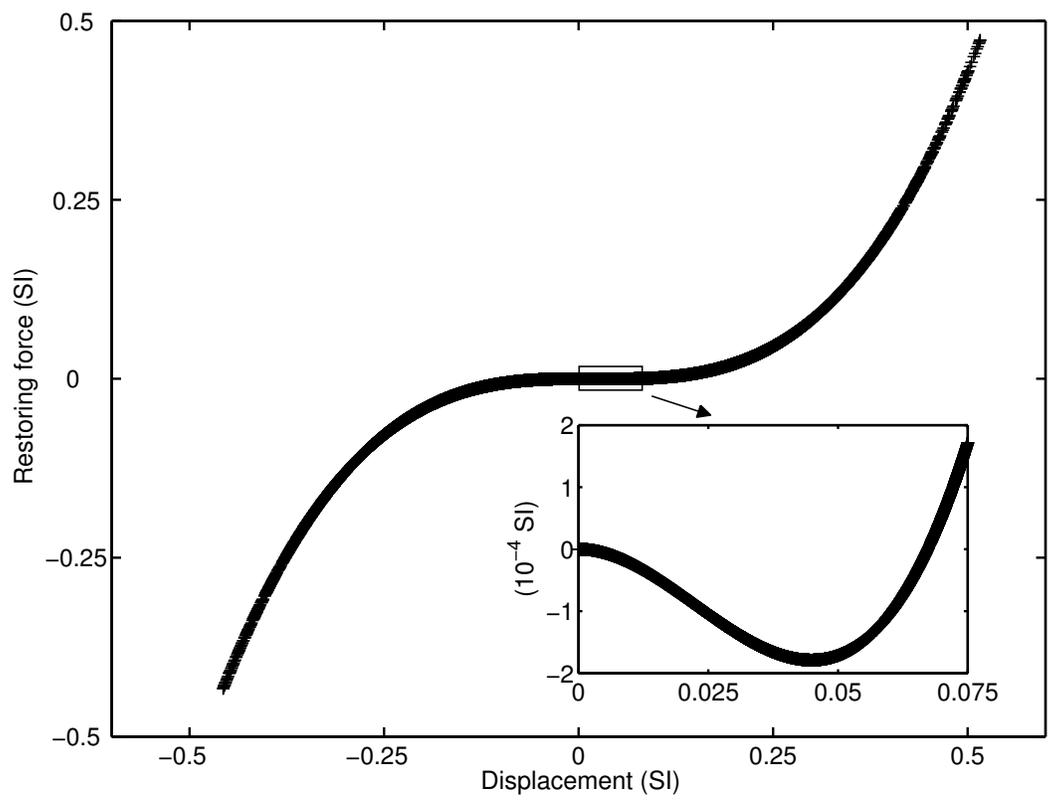} \\
\caption{Silverbox nonlinear restoring force synthesised using the nonlinear coefficients averaged at 2441 $Hz$ in Table~\ref{Table:Silverbox_NLCoeff}.}
\label{Fig:Silverbox_NL}
\end{center}
\end{figure}

The real and imaginary parts of the estimated $c_{1}$ and $c_{2}$ are depicted versus frequency in Fig.~\ref{Fig:Silverbox_NLCoeff}. They are presented for two distinct sampling frequencies, namely the original rate of 2441 $Hz$ (in black) and a five-times greater rate of 12205 $Hz$ (in blue). The imaginary parts of the coefficients are seen to be more than one order of magnitude smaller than the real parts. The frequency dependence of the real parts is also found to be reduced when increasing the sampling frequency, as a result of the decrease of the error inherent to the discrete- to continuous-time conversion~[\cite{Schoukens_Book,Relan_DiscrCont}]. Using the averaged values of the coefficients calculated at 2441 $Hz$ (doing so at 12205 $Hz$ leads to very similar values, see Table~\ref{Table:Silverbox_NLCoeff}), Fig~\ref{Fig:Silverbox_NL} displays the synthesis of the nonlinear restoring force $c_{1} \: y^{2}_{nl}(t) + c_{2} \: y^{3}_{nl}(t)$ in the system. The cubic nonlinearity dominates this force-displacement curve, even though it additionally exhibits a slight asymmetry due to the quadratic nonlinearity, particularly visible in the close-up plot.

\begin{table}[p]
\centering
\begin{tabular*}{0.75\textwidth}{@{\extracolsep{\fill}} c c c}
\hline
Sampling frequency ($Hz$) & $c_{1}$ (SI) & $c_{2}$ (SI) \\
\hline
2441 & -0.256 & 3.98 \\
12205 & -0.267 & 3.96 \\
\hline
\end{tabular*}
\caption{Spectral average over the input band of the real parts of the nonlinear coefficients $c_{1}$ and $c_{2}$ for sampling frequencies of 2441 and 12205 $Hz$.}
\label{Table:Silverbox_NLCoeff}
\end{table} 

\FloatBarrier
\subsection{Statistical analysis over multiple input realisations}\label{Sec:Statistics}

In this section, a statistical analysis of the Silverbox identification is conducted. Table~\ref{Table:Silverbox_Statistics} lists the means and standard deviations of the 13 grey-box model parameters calculated over 50 input realisations, \textit{i.e.} 50 multisine input signals with different phase spectra independently drawn from a uniform distribution on $\left[0,\:2\pi\right)$~[\cite{Schoukens_Book}]. One first observes in this table the reasonably low variability of all estimates. Compared to parameters in $\left( \mathbf{B},\mathbf{C},\mathbf{D},\mathbf{E} \right)$, the linear dynamic parameters in matrix $\mathbf{A}$ are also found to exhibit a ratio between standard deviation and mean values lower by one order of magnitude.\\

\begin{figure}[ht]
\begin{center}
\begin{tabular}{c c}
\subfloat[]{\label{Fig:Silverbox_QuadNL_Real_Stat}\includegraphics[height=56mm]{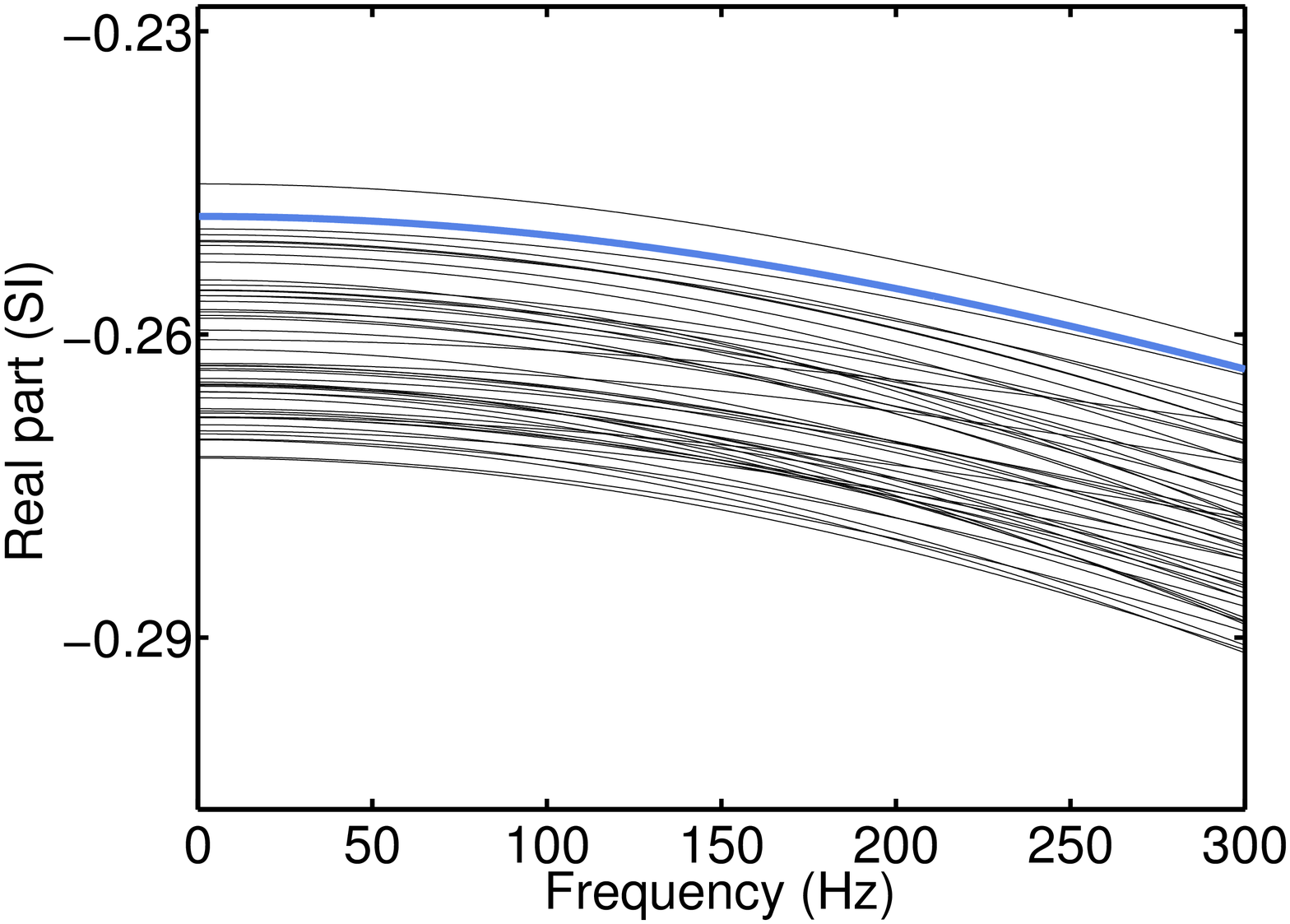}} &
\subfloat[]{\label{Fig:Silverbox_CubicNL_Real_Stat}\includegraphics[height=56mm]{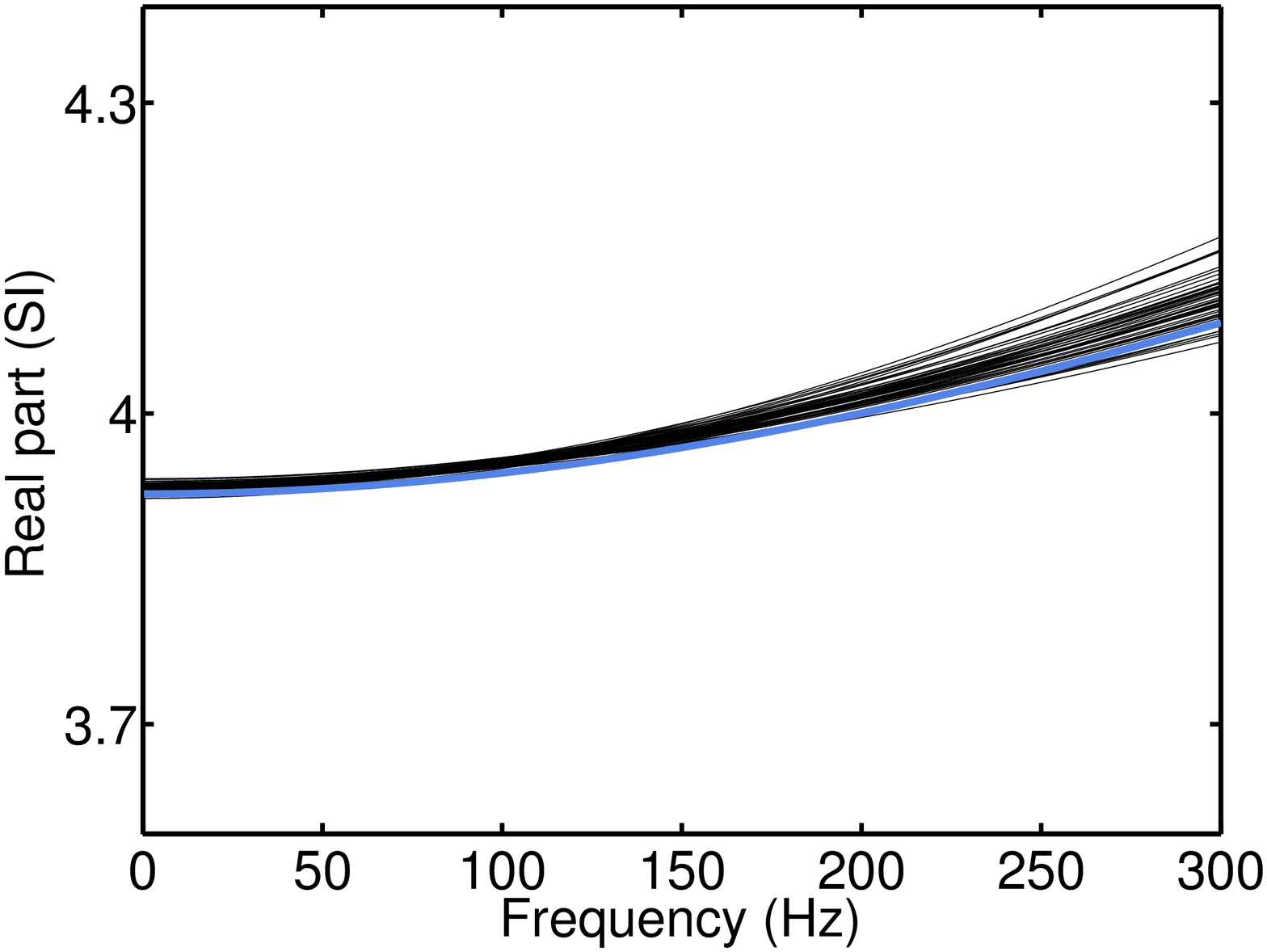}} \\
\end{tabular}
\caption{Variability of the frequency-dependent nonlinear coefficients (a) $c_{1}$ and (b) $c_{2}$ over 50 input realisations. In the two plots, the limits of the vertical axes correspond to $\pm \; 10\%$ of the coefficient mean values. In blue, the two coefficients obtained in Section~\ref{Sec:NLCoeff} are reproduced.}
\label{Fig:Silverbox_NLCoeff_Stat}
\end{center}
\end{figure}

In Table~\ref{Table:Silverbox_ModalStatistics}, similar statistics are given for the modal and physical parameters of the Silverbox system. The variability in the estimation of the linear natural frequency and damping ratio is marginal, although damping ratio estimates are seen to be comparatively more scattered. Similarly, the physical coefficient $c_{1}$ of the quadratic nonlinearity features a standard-deviation-over-mean ratio of -1.81 $\%$, compared to 0.18 $\%$ for the cubic coefficient. This is studied in Fig.~\ref{Fig:Silverbox_NLCoeff_Stat}, where the 50 frequency-dependent estimates of $c_{1}$ and $c_{2}$ are depicted. In the two plots, the limits of the vertical axes correspond to $\pm \; 10\%$ of the coefficient mean values, confirming the greater variability of $c_{1}$.\\

\vspace*{0.5cm}\begin{table}[ht]
\centering
\begin{tabular*}{0.85\textwidth}{@{\extracolsep{\fill}} c c c c}
\hline
Parameter & Mean (SI) & Standard deviation ($\times 100$) & Std. dev./mean ($\%$) \\
\hline
$A(1,1)$ & 0.97 & 0.47 & 0.49 \\
$A(2,1)$ &-0.16 & 0.10 &-0.59 \\
$A(1,2)$ & 0.19 & 0.11 & 0.58 \\
\vspace*{0.1cm}$A(2,2)$ & 0.98 & 0.47 & 0.49 \\
$B(1,1)$ &-0.05 & 0.41 &-7.88 \\
\vspace*{0.1cm}$B(2,1)$ &-0.14 & 0.50 &-3.66 \\
$C(1,1)$ &-1.12 & 3.26 &-2.90 \\
\vspace*{0.1cm}$C(1,2)$ & 0.20 & 3.75 &18.80 \\
\vspace*{0.1cm}$D(1,1)$ & 0.003& 0.04 &13.47 \\
$E(1,1)$ &-0.01 & 0.11 &-8.34 \\
$E(2,1)$ &-0.04 & 0.15 &-4.25 \\
$E(1,2)$ & 0.20 & 1.60 & 7.83 \\
$E(2,2)$ & 0.54 & 2.01 & 3.71 \\
\hline
\end{tabular*}
\caption{Statistics (means and standard deviations) of the 13 grey-box state-space parameters calculated over 50 input realisations. The ratios between standard deviation and mean values are also given in $\%$.}
\label{Table:Silverbox_Statistics}
\end{table} 

\begin{table}[ht]
\centering
\begin{tabular*}{1.00\textwidth}{@{\extracolsep{\fill}} c c c c}
\hline
Parameter & Mean & Standard deviation ($\times 100$) & Std. dev./mean ($\%$) \\
\hline
Natural frequency & 68.79 ($Hz$) & 1.00 & 0.01 \\
Damping ratio & 4.91 ($\%$) & 0.6 & 0.12 \\
$c_{1}$ & -0.26 (SI) & 0.48 & -1.81 \\
$c_{2}$ & 3.99 (SI) & 0.72 & 0.18 \\
\hline
\end{tabular*}
\caption{Statistics of the Silverbox modal and physical parameters calculated over 50 input realisations.}
\label{Table:Silverbox_ModalStatistics}
\end{table}

\FloatBarrier

\begin{figure}[p]
\begin{center}
\includegraphics[width=140mm]{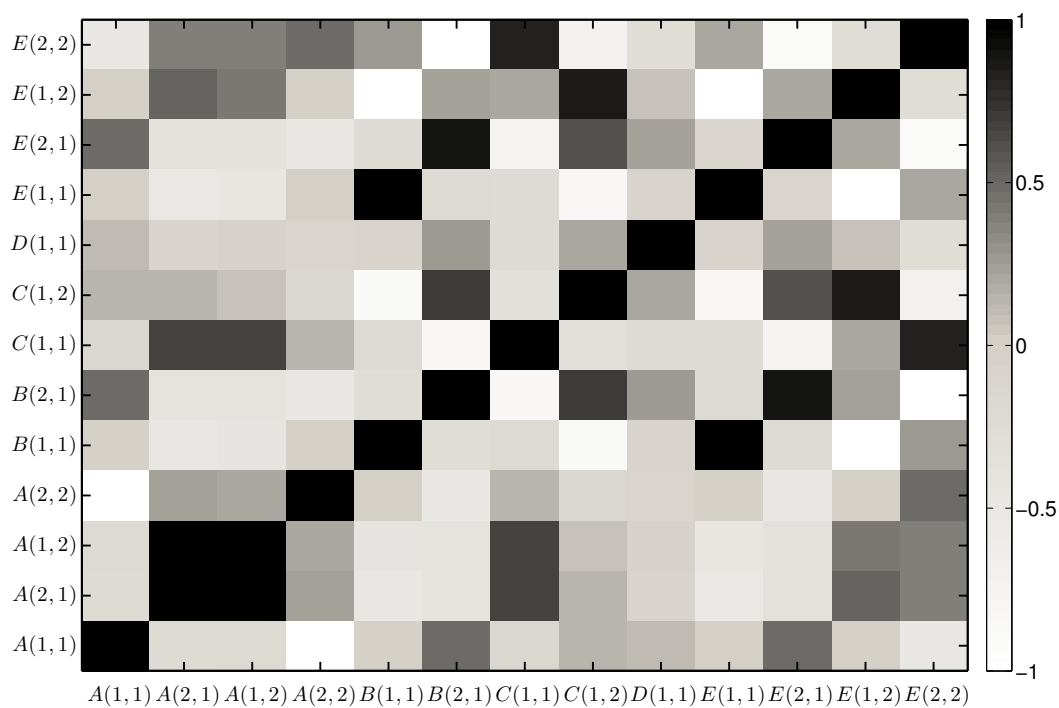} \\
\caption{Correlation matrix of the 13 grey-box state-space parameters calculated over 50 input realisations.}
\label{Fig:Silverbox_Correlation}
\end{center}
\end{figure}

\begin{figure}[p]
\begin{center}
\begin{tabular}{c c}
\subfloat[]{\label{Fig:Silverbox_Corr_AA}\includegraphics[height=58mm]{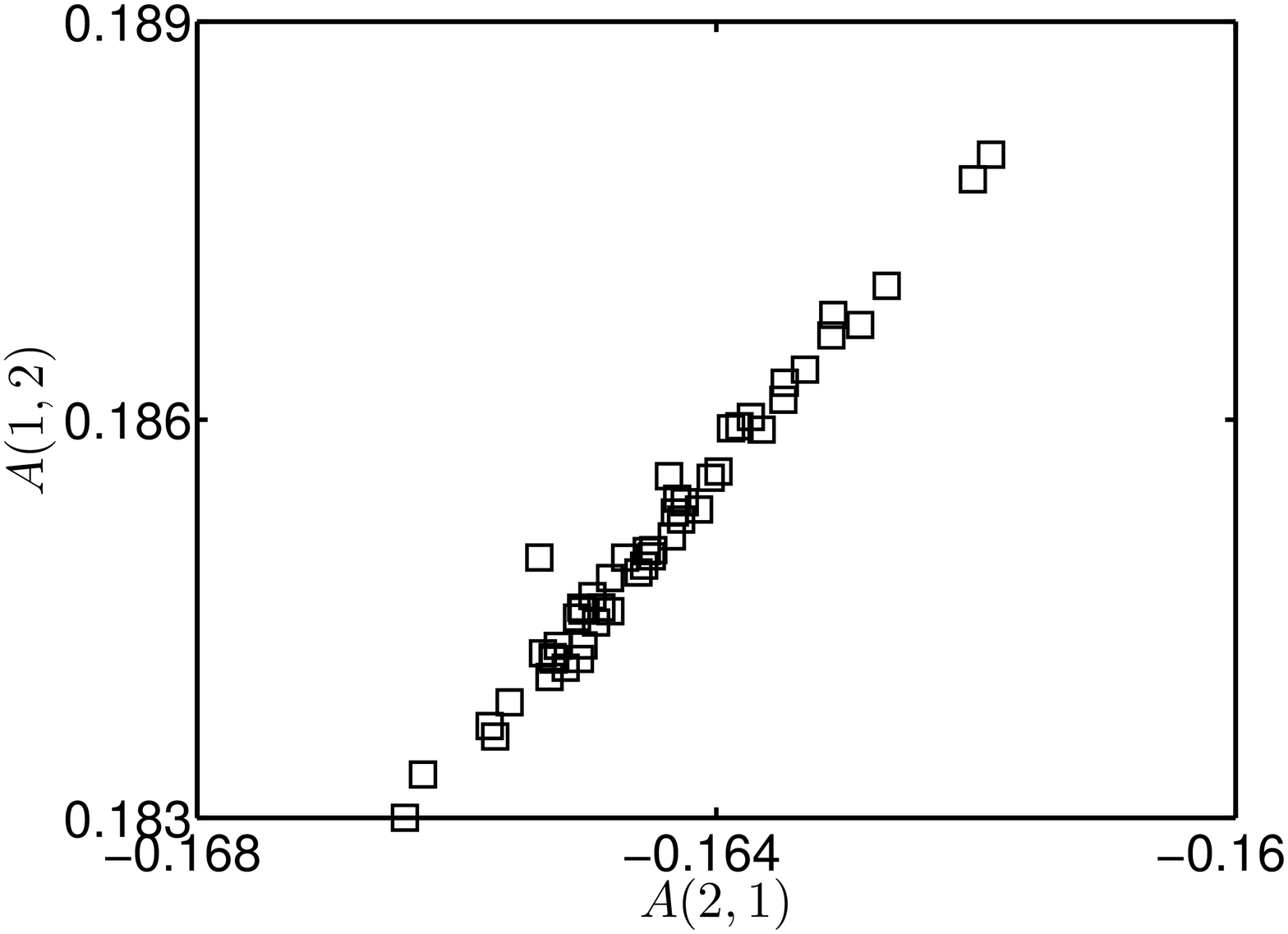}} &
\subfloat[]{\label{Fig:Silverbox_Corr_BE}\includegraphics[height=58mm]{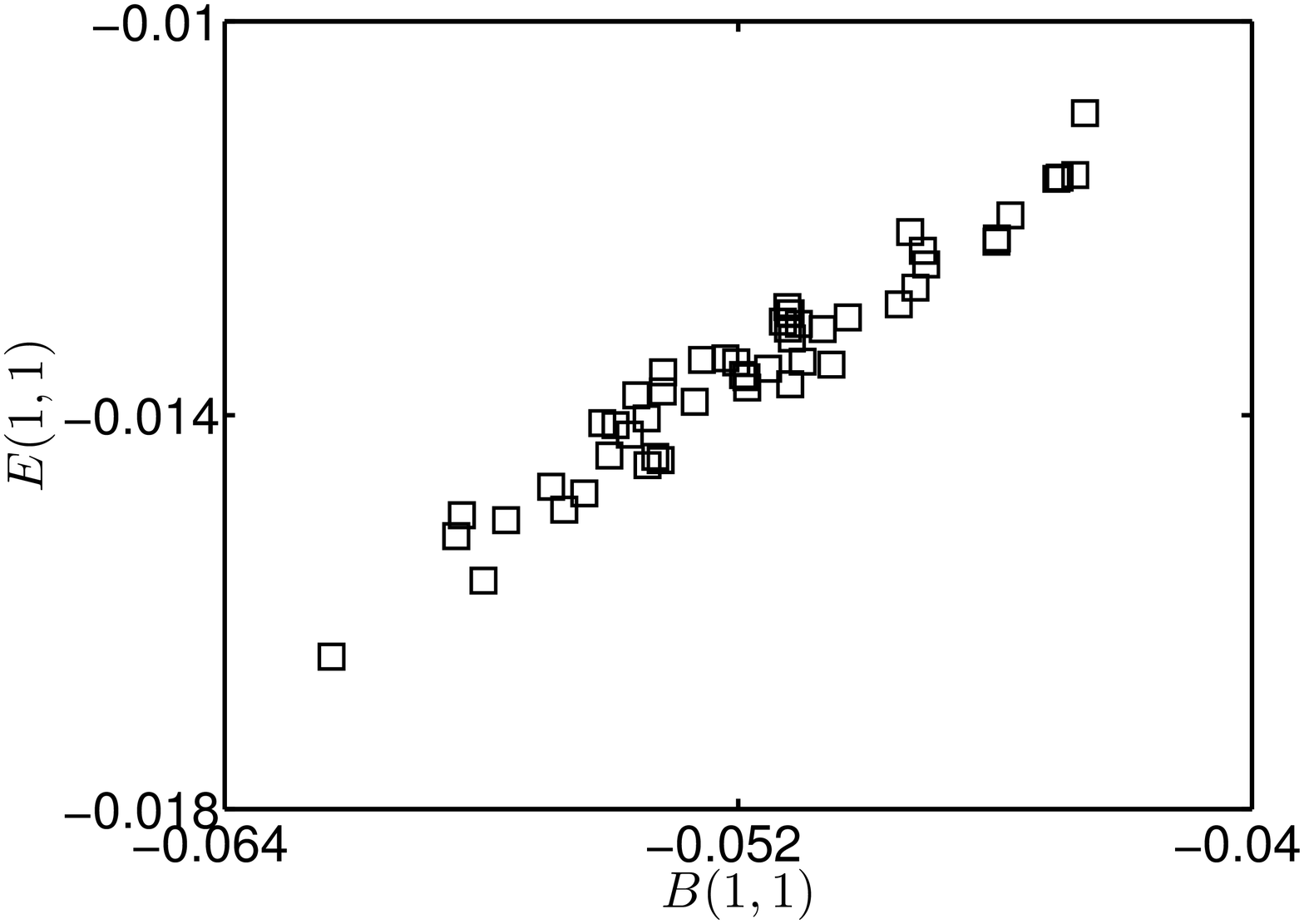}} \\
\end{tabular}
\caption{Evidence of the correlation existing between (a) the $A(2,1)$ and $A(1,2)$ parameters and (b) the $B(1,1)$ and $E(1,1)$ parameters.}
\label{Fig:Silverbox_CorrExamples}
\end{center}
\end{figure}

The existence of correlation between the estimated state-space parameters is inspected in Fig.~\ref{Fig:Silverbox_Correlation}. This figure presents the correlation matrix of the parameters, computed by normalising the rows and columns of their covariance matrix by the associated standard deviations. Multiple off-diagonal elements with significant magnitudes are observed, with 5 elements in the upper triangular block larger than 0.8. This is illustrated in Fig.~\ref{Fig:Silverbox_CorrExamples}~(a) and~(b) by plotting the pairs of strongly correlated parameter estimates $\left\{ A(2,1),A(1,2) \right\}$ (corr. = 0.99) and $\left\{ B(1,1),E(1,1) \right\}$ (corr. = 0.97), respectively. The evidence of correlation implies that the derived state-space model is an overparametrised representation of the input-output relationship. In fact, Eq.~(\ref{Eq:Duffing}) indicates that 5 parameters in physical space are sufficient to describe the Silverbox dynamics, in comparison to the 13 grey-box parameters.\\

It is a user decision to opt for a physical- or a state-space nonlinear modelling approach. Overparametrisation in state space is expected to lead to a greater variability of the estimated parameters compared to a minimal system representation in physical space. However, in a grey-box formulation, the existence of correlation between state-space parameters results in physically interpretable parameters, namely linear modal properties and nonlinear coefficients, with little variability (see Table~\ref{Table:Silverbox_ModalStatistics}). Moreover, by increasing the model flexibility, overparametrising plays a key role in the decrease of the input-output simulation error (see Fig.~\ref{Fig:Silverbox_Error}).

\FloatBarrier
\newpage
\begin{figure}[ht]
\subsection{Comparison with other nonlinear model structures}

The grey-box state-space model proposed in Section~\ref{Sec:Model} is compared herein to a number of other nonlinear model structures. To this end, a benchmark data set measured on the Silverbox system, and described in Ref.~[\cite{Marconato_Silverbox}], is utilised. It consists of ten realisations of a multisine signal serving as estimation data, and of a filtered Gaussian noise sequence with increasing RMS value functioning as test data. In Fig.~\ref{Fig:Silverbox_Comparison}, the RMS error evaluated on test data is plotted versus number of parameters for 10 different nonlinear models derived based on estimation data, namely (1) linear, (2) Hammerstein, (3) nonlinear feedback, (4) linear fractional representation, (5) locally-linear state-space, (6) black-box state-space with sigmoidal nonlinearities, (7) black-box state-space with polynomial nonlinearities, (8) support vector machines, (9) neural network, and (10) grey-box state-space models. Detailed information about the features of the different models is to be found in Ref.~[\cite{Marconato_Silverbox}] and the references therein. The model structure introduced in the present paper (number (10), in blue) is seen to achieve a good trade-off between accuracy and parsimony. It shows a slightly larger RMS error than black-box state-space models (numbers (6 -- 7)), owing to the existence of an extrapolation region in the test data sequence where it fails to predict the Silverbox response accurately. However, it possesses the lowest number of parameters compared to other nonlinear state-space models (numbers (5 -- 6 -- 7)). It should finally be noted that models (2 -- 3 -- 4) are block-oriented models, which perform very well in the case of a SISO system with a single resonance but, unlike state-space models, generalise limitedly to larger-scale systems with multiple resonances and nonlinearities.

\vspace*{0.3cm}\begin{center}
\begin{overpic}[width=140mm,tics=50]
{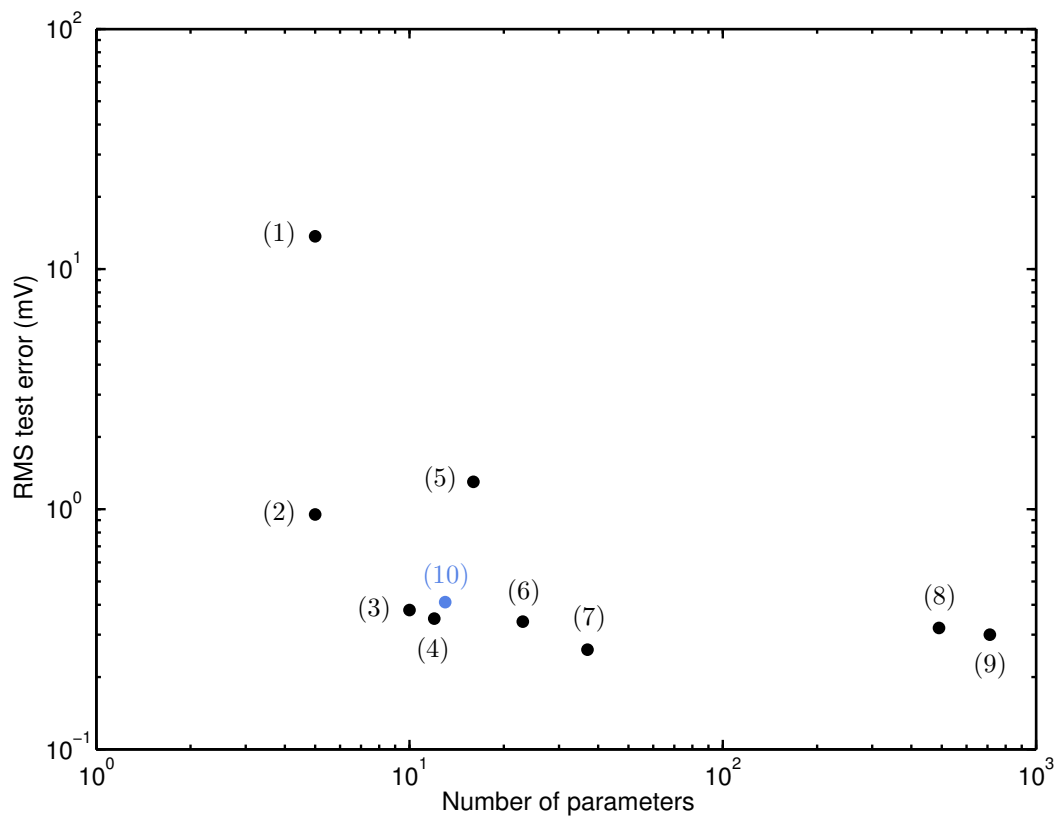}
\put(242,550){\small (1)}
\put(242,285){\small (2)}
\put(332,194){\small (3)}
\put(388,155){\small (4)}
\put(395,317){\small (5)}
\put(474,210){\small (6)}
\put(536,185){\small (7)}
\put(869,205){\small (8)}
\put(917,140){\small (9)}
\definecolor{myblue}{RGB}{85,130,230}
\put(394,225){\small \textcolor{myblue}{(10)}}
\end{overpic}
\caption{Comparison of 10 different nonlinear model structures based on benchmark data, namely (1) linear, (2) Hammerstein, (3) nonlinear feedback, (4) linear fractional representation, (5) locally-linear state-space, (6) black-box state-space with sigmoidal nonlinearities, (7) black-box state-space with polynomial nonlinearities, (8) support vector machines, (9) neural network, and (10) grey-box state-space (this paper, in blue) models. Detailed information about the features of the different models is to be found in Ref.~[\cite{Marconato_Silverbox}] and the references therein.}
\label{Fig:Silverbox_Comparison}
\end{center}
\end{figure}

\FloatBarrier

\newpage
\section{Experimental demonstration on a nonlinear beam benchmark}~\label{Sec:NLBeam}

\begin{figure}[ht]
\begin{center}
\includegraphics[width=100mm]{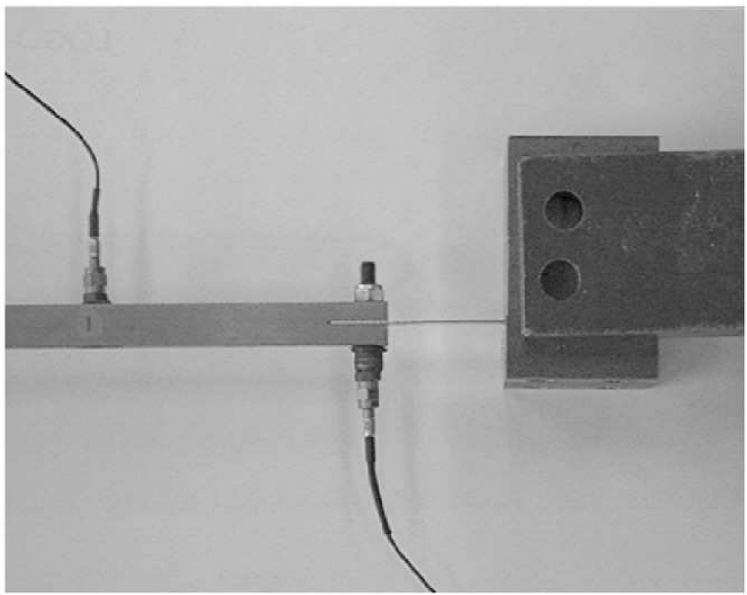}
\caption{Close-up picture of the connection between the main linear beam and the thin beam (top view).} 
\label{Fig:NLBeam}
\end{center}
\end{figure}

This section demonstrates the identification procedure of Section~\ref{Sec:ID} using the \textit{Ecole Centrale de Lyon} (ECL) benchmark structure~[\cite{Thouverez_ECL}]. It is an experimental system comprising a thin beam behaving as a localised nonlinear stiffness component, connected to a linear thick beam with well-separated, lowly-damped modes (see the geometric properties in Table~\ref{Table:NLBeam_properties}). The system is clamped on both sides and is entirely made up of steel. A close-up picture of the connection between the main linear beam and the thin beam is displayed in Fig.~\ref{Fig:NLBeam}. The nature of the nonlinearity in the system is geometric, and is physically related to the substantially large displacements experienced by the thin beam at high forcing amplitude compared to its thickness.\\

\begin{table}[ht]
\centering
\begin{tabular*}{0.75\textwidth}{@{\extracolsep{\fill}} c c c c}
\hline
Component & Length ($m$) & Width ($mm$) & Thickness ($mm$) \\
\hline
Main beam & 0.7 & 14 & 14 \\
Thin beam & 0.04 & 14 & 0.5 \\
\hline
\end{tabular*}
\caption{Geometric properties of the nonlinear beam benchmark.}
\label{Table:NLBeam_properties}
\end{table} 

The system was instrumented using 7 accelerometers, regularly spaced along the main beam. Excitation signals were applied in a horizontal plane using a shaker connected to the main beam through a stinger attached 20 $cm$ away from the clamping. An impedance head was used to measure force and acceleration signals at the excitation location. Random phase multisine inputs were generated, considering a sampling frequency of 1600 $Hz$ and a frequency resolution of about 0.2 $Hz$ (the number of time samples was 8192). The selected bandwidth of interest ranges from 20 to 100 $Hz$ excluding DC and, in each experiment, 10 periods of input-output data were collected. The first 2 periods in each sequence were rejected to achieve steady state, and the final period was saved for validation. Because of dynamic interactions existing between the shaker and the structure, the input processed herein is the voltage signal measured at the output of the measurement setup amplifier.\\

\begin{figure}[ht]
\begin{center}
\includegraphics[width=140mm]{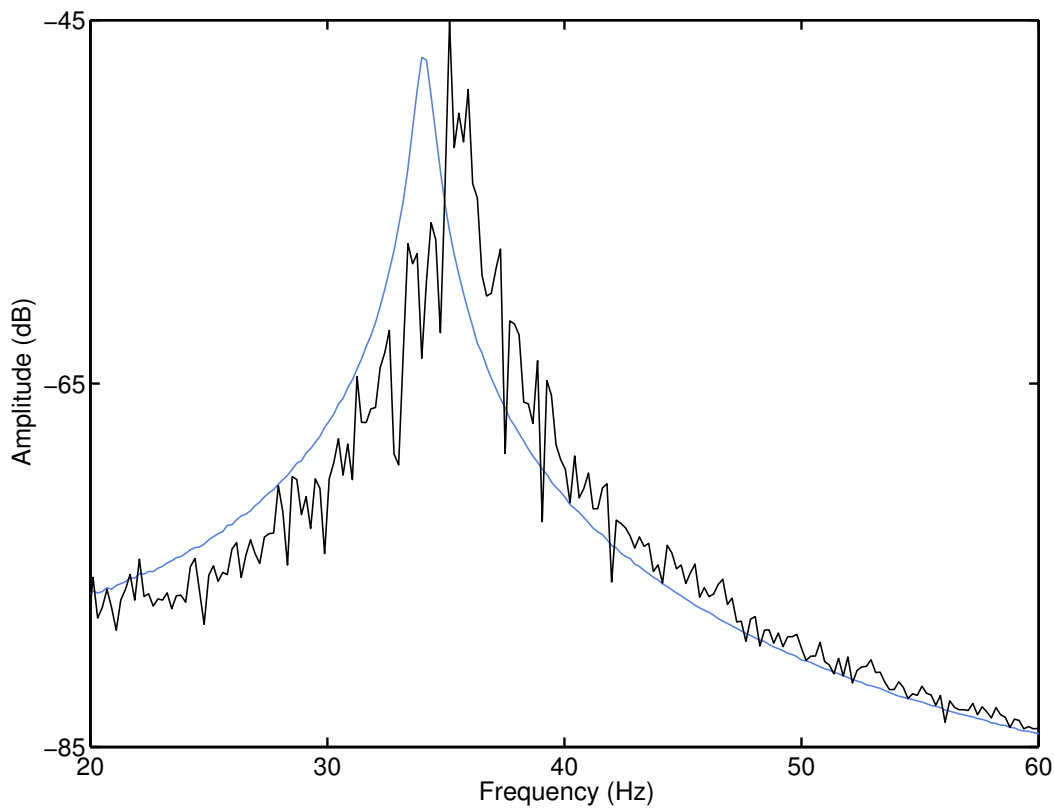} \\
\caption{Comparison of FRFs measured at the main beam tip at 0.1 (in blue) and 1.2 (in black) $V$ RMS.}
\label{Fig:ECL_FRF}
\end{center}
\end{figure}

At 0.1 $V$ RMS input amplitude (corresponding approximately to 1.2 $N$ RMS), the system exhibits linear dynamics, as expected from the geometric nature of the nonlinearity. The modal properties of the first structural mode estimated at this level are given in Table~\ref{Table:ECL}. Fig.~\ref{Fig:ECL_FRF} compares FRFs measured at the main beam tip (sensor number 7) at 0.1 $V$ RMS and 1.2 $V$ RMS (corresponding approximately to 11.1 $N$ RMS), revealing typical stochastic nonlinear distortions and a shift of the resonance frequency of about 2 $Hz$.

\begin{table}[ht]
\begin{center}
\begin{tabular}{c c}
\hline
Natural frequency ($Hz$) & Damping ratio ($\%$) \\
34.08 & 1.11 \\
\hline
\end{tabular}
\caption{Linear natural frequency and damping ratio of the first structural mode of the nonlinear beam benchmark estimated at 0.1 $V$ RMS.} 
\label{Table:ECL}
\end{center}
\end{table}

\begin{figure}[p]
\begin{center}
\begin{tabular}{c}
\subfloat[]{\label{Fig:ECL_FDError_4}\includegraphics[width=140mm]{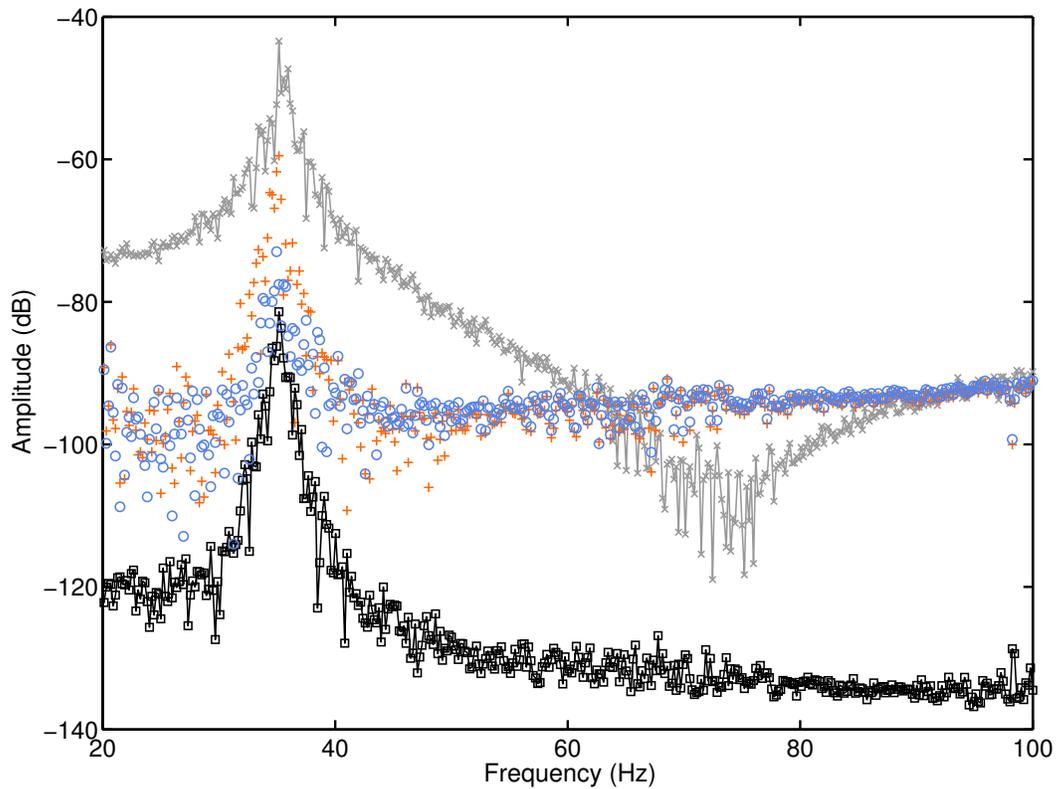}}\\
\subfloat[]{\label{Fig:ECL_FDError_7}\includegraphics[width=140mm]{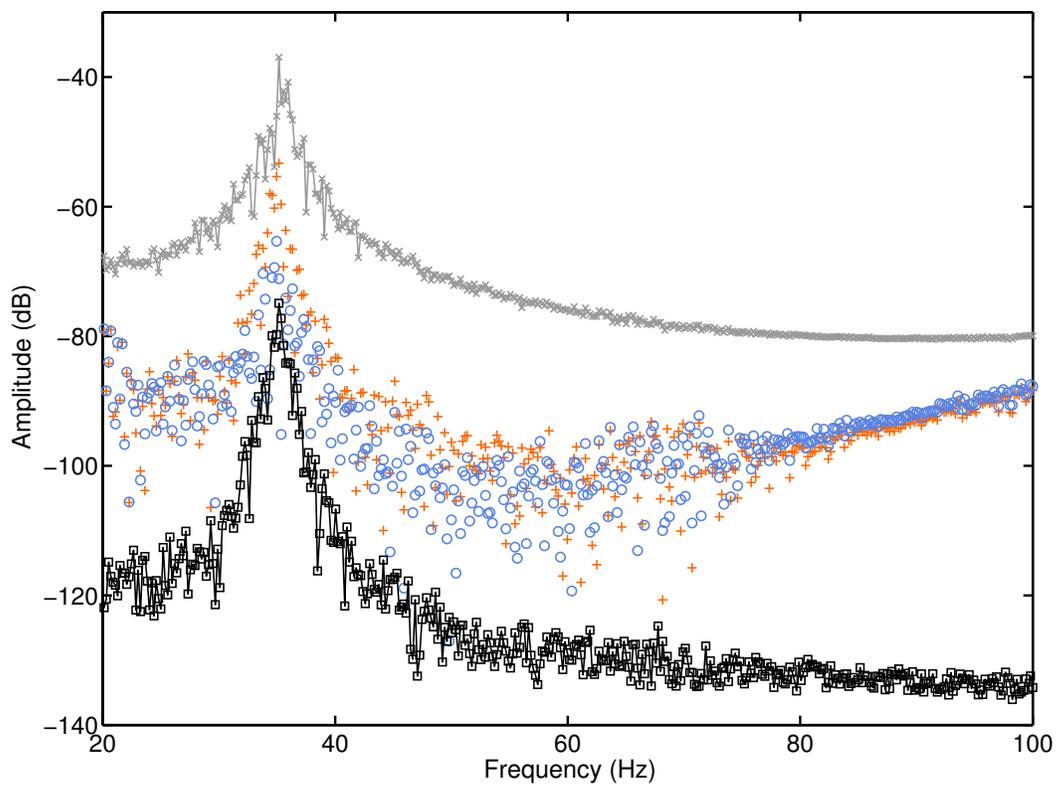}}\\
\end{tabular}
\caption{Frequency-domain behaviour of the validation model error over the 20 -- 100 $Hz$ band, featuring the output spectrum (in grey), initial (in orange) and final (in blue) grey-box state-space error levels, and noise level (in black). (a) Sensor 4 and (b) sensor 7.}
\label{Fig:ECL_FDError}
\end{center}
\end{figure}

\begin{figure}[ht]
\begin{center}
\includegraphics[width=140mm]{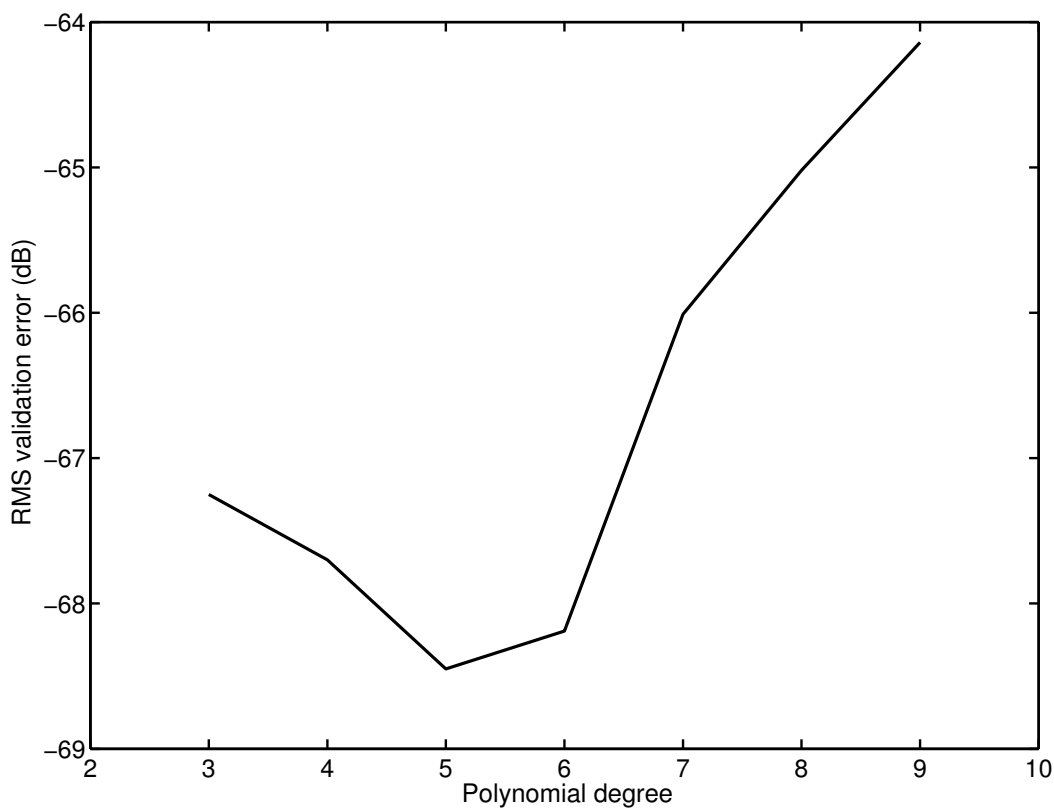} \\
\caption{Time-domain RMS validation error for grey-box models with increasing polynomial degree.}
\label{Fig:ECLBeam_Degree}
\end{center}
\end{figure}

A grey-box model of the beam dynamics is constructed at 1.2 $V$ RMS input amplitude considering nonlinear terms in the state equation formed using the displacement measured at the main beam tip (sensor number 7), \textit{i.e.} the displacement measured at the nonlinearity location, noted $y_{nl} = y_{7}$. A fifth-degree polynomial in $y_{nl}$ is used to describe the geometrically nonlinear effects induced by the thin beam, leading to a second-order model in state space comprising 35 parameters. Fig.~\ref{Fig:ECL_FDError} presents validation model error plots in the frequency domain, featuring the output spectrum (in grey), initial (in orange) and final (in blue) grey-box state-space error levels, and noise level (in black). Two sensor locations are represented, namely (a) sensor 4 close to the mid-span of the main beam and (b) sensor 7 at the nonlinearity location. In the resonance region, the errors of the initial and final models are seen to lie about 15 and 30 $dB$ below the output level, respectively. Out of resonance, initial and final model errors correspond and possess an amplitude larger than the noise floor. In Fig.~\ref{Fig:ECLBeam_Degree}, the time-domain RMS validation error is plotted for grey-box models with increasing polynomial degree, showing a clear minimum corresponding to a representation of degree five.\\

\begin{figure}[p]
\begin{center}
\begin{tabular}{c}
\subfloat[]{\label{Fig:ECL_Initialisation}\includegraphics[width=140mm]{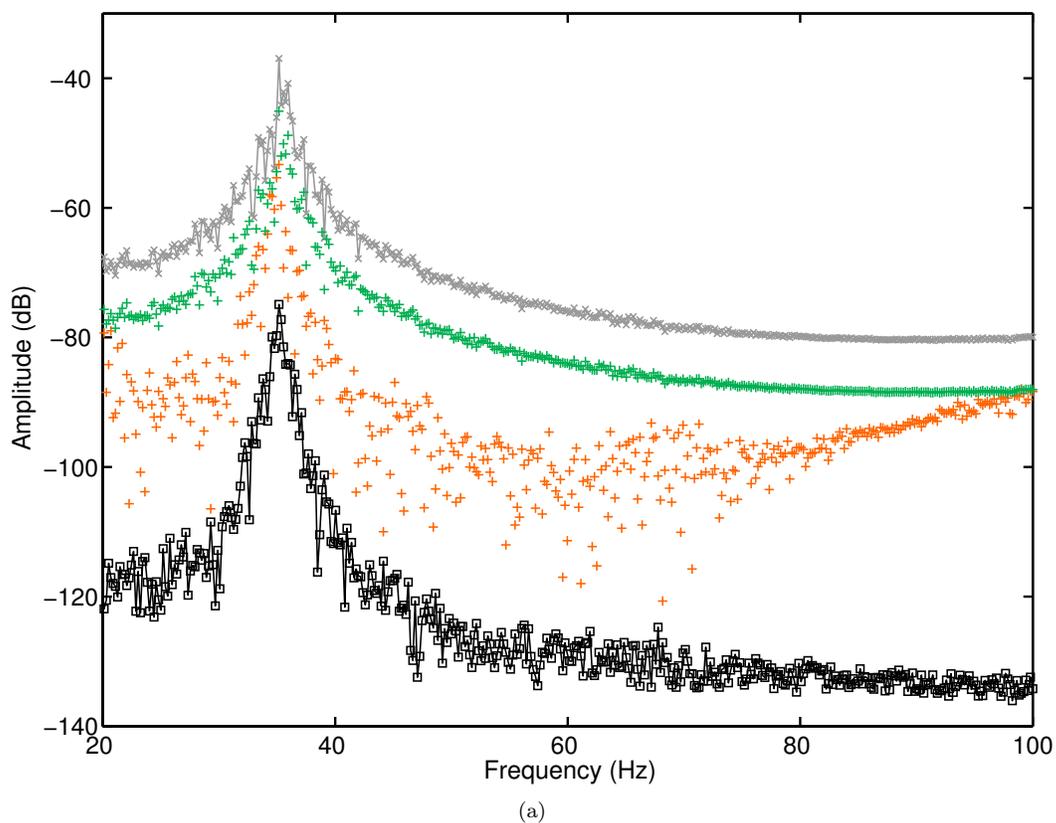}}\\
\subfloat[]{\label{Fig:ECL_GreyBlack}\includegraphics[width=140mm]{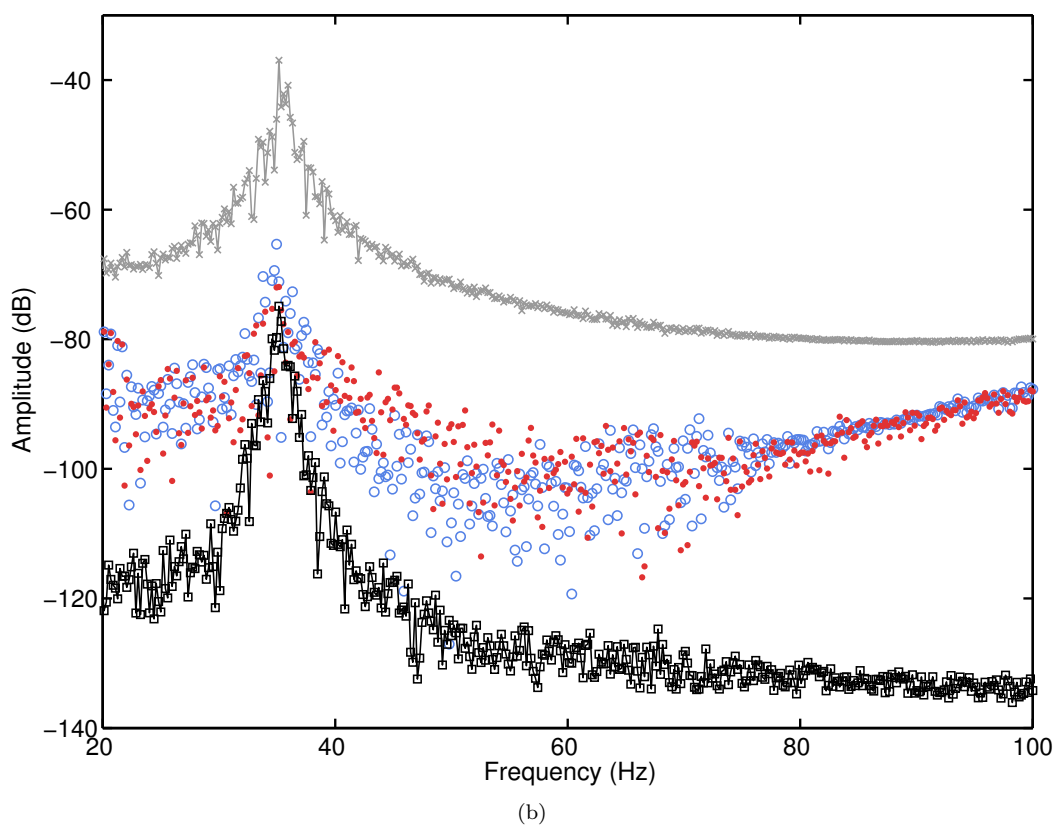}}\\
\end{tabular}
\caption{Frequency-domain comparison at sensor 7 of the grey-box and black-box state-space modelling approaches. Validation output spectra and noise levels are plotted in grey and black, respectively. (a) Initial grey-box model obtained using nonlinear subspace identification (in orange) and initial linear black-box model (in green); (b) final grey-box (in blue) and black-box (in red) models.}
\label{Fig:ECL_Compare}
\end{center}
\end{figure}

\begin{figure}[p]
\begin{center}
\includegraphics[width=140mm]{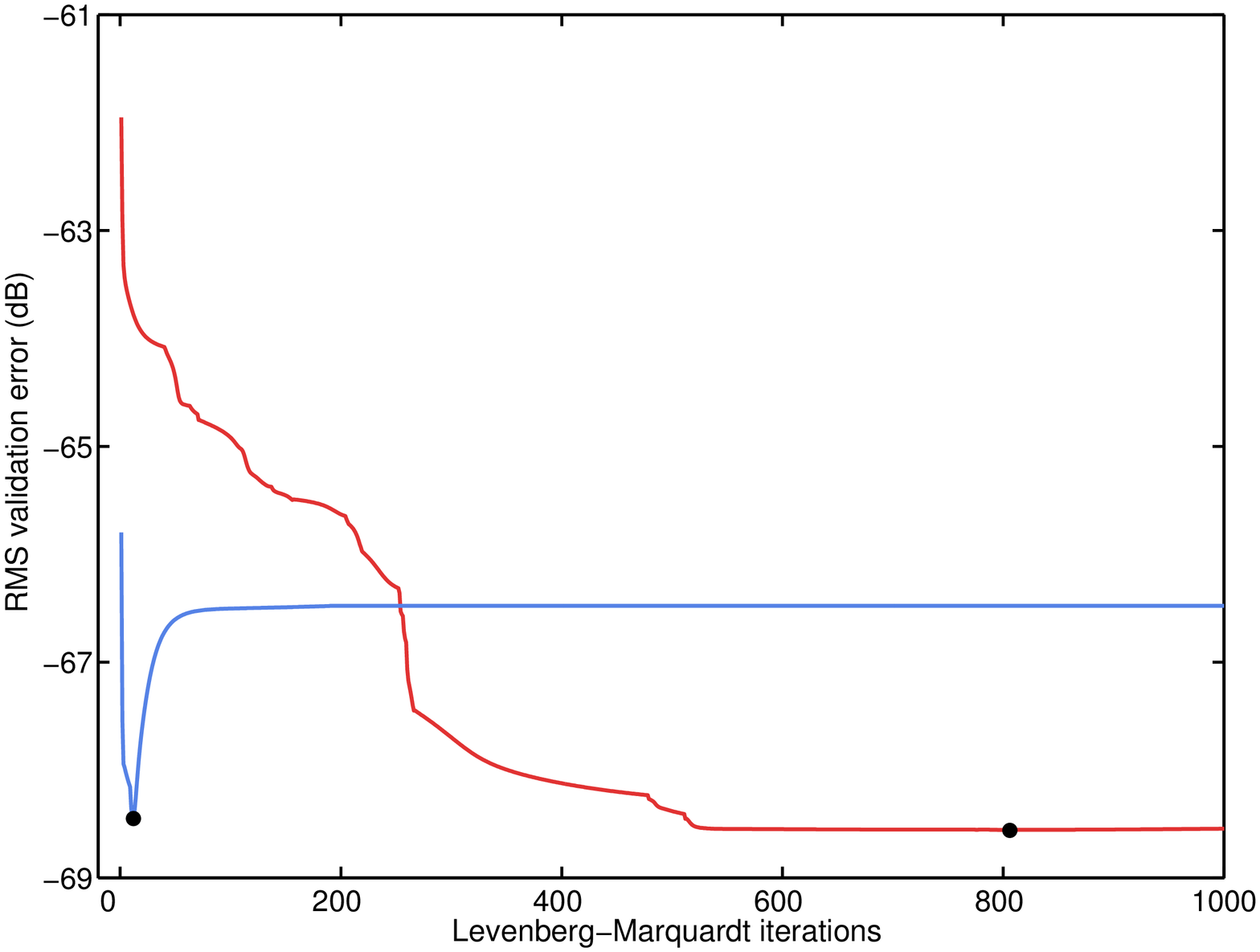} \\
\caption{Decrease of the RMS validation error over 1000 Levenberg-Marquardt iterations for the grey-box (in blue) and black-box (in red) approaches. Optimal models are located using black dots.}
\label{Fig:ECL_Iterations}
\vspace*{0.2cm}
\includegraphics[width=140mm]{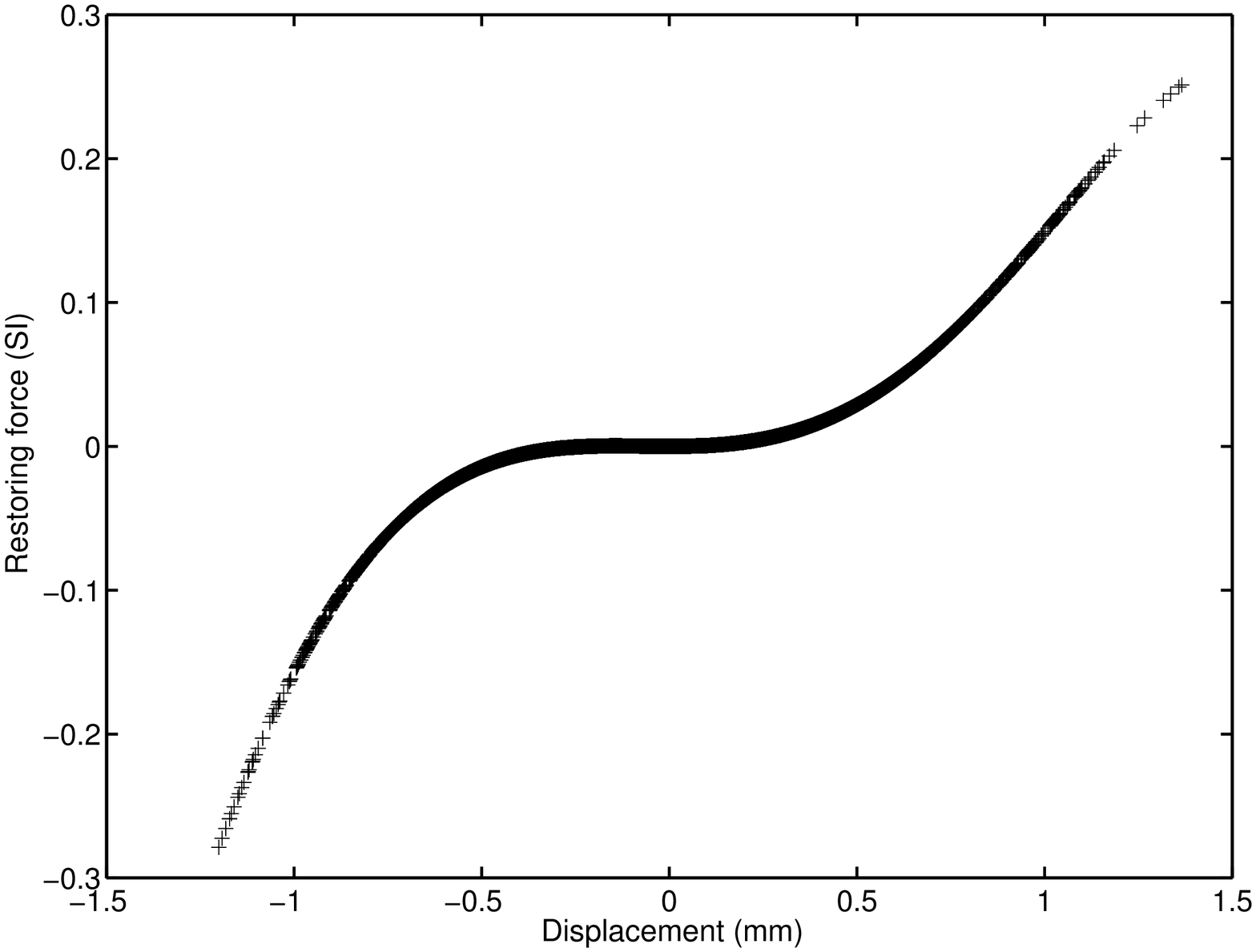} \\
\caption{Physical restoring force of the nonlinear beam benchmark synthesised using the averaged nonlinear coefficients listed in Table~\ref{Table:ECL_NLCoeff}.}
\label{Fig:ECL_NL}
\end{center}
\end{figure}

In Fig.~\ref{Fig:ECL_Compare}, the obtained fifth-degree grey-box model is compared at sensor 7 with a black-box model including in the state equation a fifth-degree multivariate polynomial involving 36 monomial combinations of the two state variables, and resulting into a state-space model with 99 parameters. Similarly to the Silverbox analysis, Fig.~\ref{Fig:ECL_Compare} shows that the initial grey-box model (in orange) provided by the nonlinear subspace identification algorithm outperforms the initial linear model (in green) of the black-box method. It is also observed that the two final models (grey-box in blue and black-box in red) achieve a comparable validation error level throughout the frequency band of interest, except in the vicinity of the system resonance, where the black-box model reaches the noise floor, 5 $dB$ below the grey-box model error level. Finally, the study in Fig.~\ref{Fig:ECL_Iterations} of the Levenberg-Marquardt iterations necessary to achieve the final estimation of the state-space parameters reveals that optimal grey-box and black-box models are reached in 12 and 806 iterations, respectively.\\

The synthesis of the physical restoring force acting at the nonlinearity location, and expressed as the four-term sum $c_{1} \: y^{2}_{nl}(t) + c_{2} \: y^{3}_{nl}(t) + c_{3} \: y^{4}_{nl}(t) + c_{4} \: y^{5}_{nl}(t)$, is performed in Fig.~\ref{Fig:ECL_NL}. The nonlinear coefficients in this expression were obtained by averaging over the 20 -- 100 $Hz$ band their frequency-dependent estimates generated as explained in Section~\ref{Sec:NLCoeff}. They are listed in Table~\ref{Table:ECL_NLCoeff}. A dominant odd behaviour is noted in Fig.~\ref{Fig:ECL_NL}, mainly associated with the cubic nonlinearity in the model. For positive displacements, an asymmetry is also remarked, visible as a softening of the force. This effect might be attributed to an imperfection in the clamping of the thin beam component, as already pointed out in Ref.~[\cite{Grappasonni_HS}].

\begin{table}[ht]
\vspace*{1cm}
\begin{center}
\begin{tabular}{c c c c}
\hline
$c_{1}$ (SI) & $c_{2}$ (SI) & $c_{3}$ (SI) & $c_{4}$ (SI) \\
$4.05 \; 10^{4}$ & $1.80 \; 10^{8}$ & $-4.46 \; 10^{10}$  & $-2.67 \; 10^{13}$ \\
\hline
\end{tabular}
\caption{Spectral average over the 20 -- 100 $Hz$ band of the real parts of the nonlinear coefficients for a sampling frequency of 1600 $Hz$.} 
\label{Table:ECL_NLCoeff}
\end{center}
\end{table}

\section{Conclusions}

The objective of the present paper was to introduce a grey-box state-space modelling framework to support the identification of nonlinear mechanical vibrations. Assuming nonlinearities localised in physical space, which is a generic case in mechanics, this framework was shown to pave the way for an important decrease in the number of parameters with respect to classical black-box state-space modelling. A general comparison of the grey-box and black-box identification features is drawn in Table~\ref{Table:Comparison}.\\

\begin{table}[ht]
\begin{center}
\begin{tabular*}{1.00\textwidth}{@{\extracolsep{\fill}} c c}
\hline
\textbf{Grey-box state-space model}  & \textbf{Black-box state-space model} \\
& \\
Nonlinear model terms are & Nonlinear model terms are \\
univariate polynomials & multivariate polynomials \\
\vspace{3mm}
with a low number of parameters. & with a high number of parameters. \\
Nonlinear coefficients have & Nonlinear coefficients have \\ 
\vspace{3mm}
physical meaning. & no direct interpretation. \\
Identification procedure includes 2 steps & Identification procedure includes 4 steps \\
\vspace{3mm}
with 1 nonlinear optimisation search. & with 2 nonlinear optimisation searches. \\
Initialisation is based & Initialisation is based \\
\vspace{3mm}
on a nonlinear subspace model. & on a linearised model. \\
\hline
\end{tabular*}
\caption{General comparison of the grey-box and black-box identification frameworks.} 
\label{Table:Comparison}
\end{center}
\end{table}

The Silverbox benchmark was considered as a first experimental case study demonstrating the derived identification procedure, combining nonlinear subspace initialisation and weighted least-squares optimisation. Compared with nine other model structures, the proposed grey-box approach was shown to lead to a good compromise between fitting flexibility and parsimony. A nonlinear beam benchmark was treated as a second experimental case study, confirming the findings of the Silverbox analysis. Future research prospects include the application of the developed framework to multiple-mode and multiple-nonlinearity mechanical systems. The calculation of confidence bounds on the model parameters would be another major advance in the topic. 

\FloatBarrier
\section*{Acknowledgements}

The author J.P. No\"el is a Postdoctoral Researcher of the \textit{Fonds de la Recherche Scientifique -- FNRS} which is gratefully acknowledged. This work was also supported in part by the Fund for Scientific Research (FWO-Vlaanderen), by the Flemish Government (Methusalem), by the Belgian Government through the Inter University Poles of Attraction (IAP VII) Program, and by the ERC Advanced Grant SNLSID, under contract 320378.

\bibliographystyle{apacite}
\renewcommand\bibliographytypesize{\fontsize{10}{12}\selectfont}
\bibliography{2016_GreyBoxID_JPNoel_Bib}

\newpage
\section*{Appendix 1: Frequency-domain nonlinear subspace identification (FNSI) method}

This appendix details the FNSI algorithm~[\cite{Noel_FNSI}] used in Section~\ref{Sec:Step1} to derive non-iteratively from data initial estimates of the matrices $\left(\mathbf{A},\overline{\mathbf{B}},\mathbf{C},\overline{\mathbf{D}} \right)$ in Eqs.~(\ref{Eq:FDStateSpace}).\\

The measured output spectra are first organised in a complex-valued matrix $\mathbf{Y}_{i}^{c}$ defined as
\begin{equation}
      \mathbf{Y}^{c}_{i} = \left( \begin{array}{c c c c}
                                                    \mathbf{Y}(1) & \mathbf{Y}(2) & \ldots & \mathbf{Y}(F) \\
                                                    z_{1} \: \mathbf{Y}(1) & z_{2} \: \mathbf{Y}(2) & \ldots & z_{F} \: \mathbf{Y}(F) \\
                                                    z_{1}^{2} \: \mathbf{Y}(1) & z_{2}^{2} \: \mathbf{Y}(2) & \ldots & z_{F}^{2} \: \mathbf{Y}(F) \\
                                                    \vdots \\
                                                    z_{1}^{i-1} \: \mathbf{Y}(1) & z_{2}^{i-1} \: \mathbf{Y}(2) & \ldots & z_{F}^{i-1} \: \mathbf{Y}(F) \\
                                                    \end{array} \right) \\ \in \mathbb{C}^{\: li \times F} ,
\label{Eq:Yi}
\end{equation}
where the superscript $c$ stands for \textit{complex}, and the subscript $i$ is the user-defined number of block rows in $\mathbf{Y}_{i}^{c}$. The number of processed frequency lines is noted $F$ and the number of output variables is $l$. Defining $\mathbf{\zeta} = \text{diag} \left( z_1 \: z_2 \: \ldots \: z_F \right) \in \mathbb{C}^{\: F \times F}$ and grouping frequency lines, $\mathbf{Y}_{i}^{c}$ is recast into
\begin{equation}
      \mathbf{Y}^{c}_{i} = \left( \begin{array}{c}
                                                    \mathbf{Y} \\
                                                    \mathbf{Y} \: \mathbf{\zeta} \\
                                                    \mathbf{Y} \: \mathbf{\zeta}^2 \\
                                                    \ldots \\
                                                    \mathbf{Y} \: \mathbf{\zeta}^{i-1} \\
                     \end{array} \right).\\
\label{Eq:Yi2}
\end{equation}

The matrix of the extended input spectra is similarly formed as
\begin{equation}
      \overline{\mathbf{U}}^{c}_{i} = \left( \begin{array}{c}
                                                    \overline{\mathbf{U}} \\
                                                    \overline{\mathbf{U}} \: \mathbf{\zeta} \\
                                                    \overline{\mathbf{U}} \: \mathbf{\zeta}^2 \\
                                                    \ldots \\
                                                    \overline{\mathbf{U}} \: \mathbf{\zeta}^{i-1} \\
                      \end{array} \right) \\ \in \mathbb{C}^{\: (m + s l) \: i \times F} ,
\label{Eq:Ei2}
\end{equation}
where $m$ is the number of extended input variables. Introducing the extended observability matrix
\begin{equation}
\mathbf{\Gamma}_{i} = \left( \begin{array}{c}
                                                \mathbf{C} \\
                                                \mathbf{C} \mathbf{A} \\
                                                \mathbf{C} \mathbf{A}^{2} \\
                                                \ldots \\
                                                \mathbf{C} \mathbf{A}^{i-2} \\
                                                \mathbf{C} \mathbf{A}^{i-1} \\
                    \end{array} \right) \in \mathbb{R}^{\: l i \times n_{s}} 
\label{Eq:Gammai}
\end{equation}
and the lower-block triangular Toeplitz matrix $\mathbf{\Lambda}_{i}$
\begin{equation}
      \mathbf{\Lambda}_{i} = \left( \begin{array}{c c c c c}
                                 \overline{\mathbf{D}} & \mathbf{0} & \mathbf{0} & \ldots & \mathbf{0} \\
                                 \mathbf{C} \overline{\mathbf{B}} & \overline{\mathbf{D}} & \mathbf{0} & \ldots & \mathbf{0} \\
                                 \mathbf{C} \mathbf{A} \overline{\mathbf{B}} & \mathbf{C} \overline{\mathbf{B}} & \overline{\mathbf{D}} & \ldots & \mathbf{0} \\
                                 \vdots & \vdots & \vdots & & \vdots \\
                                 \mathbf{C} \mathbf{A}^{i-2} \overline{\mathbf{B}} & \mathbf{C} \mathbf{A}^{i-3} \overline{\mathbf{B}} & \mathbf{C} \mathbf{A}^{i-4} \overline{\mathbf{B}} & \ldots & \overline{\mathbf{D}} \\
                                 \end{array} \right)\ \in \mathbb{R}^{\: l i \times (m + s l) \: i} , 
\label{Hi}
\end{equation}
recursive substitution of the second relation into the first relation of Eqs.~(\ref{Eq:FDStateSpace}) results in the output-state-input relationship
\begin{equation}
\mathbf{Y}^{c}_{i} = \mathbf{\Gamma}_{i} \: \mathbf{X}^{c} + \mathbf{\Lambda}_{i} \: \overline{\mathbf{U}}^{c}_{i} ,
\label{Eq:OSIc}
\end{equation}
where $\mathbf{X}^{c} \in \mathbb{C}^{\: n_{s} \times F}$ is the state spectrum. To force the identified state-space model $(\mathbf{A},\overline{\mathbf{B}},\mathbf{C},\overline{\mathbf{D}})$ to be real-valued, Eq.~(\ref{Eq:OSIc}) is finally converted into the real equation
\begin{equation}
\mathbf{Y}_{i} = \mathbf{\Gamma}_{i} \: \mathbf{X} + \mathbf{\Lambda}_{i} \: \overline{\mathbf{U}}_{i}
\label{Eq:OSI}
\end{equation}
by separating the real and imaginary parts of $\mathbf{Y}_{i}^{c}$, $\mathbf{X}^{c}$ and $\overline{\mathbf{U}}_{i}^{c}$, for instance, 
\begin{equation}
    \mathbf{Y}_{i} = \left[ \mathcal{R} (\mathbf{Y}^{c}_{i}) \ \mathcal{I} (\mathbf{Y}^{c}_{i})  \right] \in \mathbb{R}^{\: li \times 2F} ,
\label{Eq:Cpx2Real}
\end{equation}
where $\mathcal{R}$ and $\mathcal{I}$ denote the real and imaginary parts, respectively.\\

Under the assumptions discussed in Ref.~[\cite{Noel_FNSI}], the orthogonal projection $\mathbf{O}_{i} = \mathbf{Y}_{i} / \overline{\mathbf{U}}^{\perp}_{i}$ and its singular value decomposition $ \mathbf{O}_{i} = \mathbf{L} \: \mathbf{S} \: \mathbf{R}^{T}$ are computed. An estimate of the extended observability matrix $\mathbf{\Gamma}_{i}$ can then be shown to be given by
\begin{equation}
	\widehat{\mathbf{\Gamma}_{i}} = \mathbf{L}_{1} \: \mathbf{S}^{1/2}_{1} ,
\label{Eq:EstGamma}
\end{equation}
where $\mathbf{L}_1$ and $\mathbf{S}_1$ contain the first $n_{s}$ left singular vectors and singular values of $\mathbf{O}_{i}$, respectively, and where hat symbols signal estimated quantities. From Eq.~(\ref{Eq:EstGamma}), $\widehat{\mathbf{C}}$ is merely extracted as the $l$ first rows of $\widehat{\mathbf{\Gamma}_{i}}$. An estimate of $\mathbf{A}$ can be calculated by exploiting the shift property \begin{equation}
\underline{\mathbf{\Gamma}_{i}} \: \mathbf{A} = \overline{\mathbf{\Gamma}_{i}} ,
\label{Eq:ShiftGamma}
\end{equation}
where $\underline{\mathbf{\Gamma}_{i}}$ and $\overline{\mathbf{\Gamma}_{i}}$ are the matrix $\widehat{\mathbf{\Gamma}_{i}}$ without its last and first $l$ rows, respectively. Matrix $\mathbf{A}$ is therefore found as the least-squares solution
\begin{equation}
\widehat{\mathbf{A}} = \underline{\mathbf{\Gamma}_{i}}^{\dagger} \: \overline{\mathbf{\Gamma}_{i}} ,
\label{Eq:EstA}
\end{equation}
where the symbol $\dagger$ denotes the pseudo-inverse operation.\\

Given estimates of $\mathbf{A}$ and $\mathbf{C}$, matrices $\overline{\mathbf{B}}$ and $\overline{\mathbf{D}}$ are finally obtained by defining the transfer function matrix $\mathbf{G}_{s}$ associated with Eqs.~(\ref{Eq:FDStateSpace}) as
\begin{equation}
\mathbf{G}_{s}(k) = \widehat{\mathbf{C}} \left( z_{k} \: \mathbf{I}^{\: n_{s} \times n_{s}} - \widehat{\mathbf{A}} \right)^{-1} \overline{\mathbf{B}} + \overline{\mathbf{D}} ,
\label{Eq:G_StateSpace}
\end{equation}
and by minimising the difference between the measured and modelled output spectra in a linear least-squares sense, \textit{i.e.} 
\begin{equation}
\widehat{\overline{\mathbf{B}}},\widehat{\overline{\mathbf{D}}} = \text{arg}\:\min_{\overline{\mathbf{B}},\overline{\mathbf{D}}} \displaystyle \sum^{F}_{k=1} \left| \mathbf{Y}(k) - \mathbf{G}_{s}(k) \overline{\mathbf{U}}(k) \right|^{2} .
\label{Eq:EstBD}
\end{equation}

\newpage
\section*{Appendix 2: Analytical calculation of the Jacobian matrix}

One first focuses on the determination of the element $J_{A_{ij}}(t) \in \mathbb{R}^{\: l}$ of the time-domain Jacobian defined as
\begin{equation}
J_{A_{ij}}(t) = \dfrac{\partial \mathbf{y}(t)}{\partial A_{i \: j}} .
\label{Eq:JAij_Def}
\end{equation}

The derivative of the output relation in Eqs.~(\ref{Eq:GreyBoxStateSpace}) with respect to $A_{i \: j}$ is given by
\begin{equation}
\begin{array}{r c l}
    \dfrac{\partial \mathbf{y}(t)}{\partial A_{i \: j}} & = & \dfrac{\partial}{\partial A_{i \: j}} \left ( \mathbf{C} \: \mathbf{x}(t) + \overline{\mathbf{D}} \: \overline{\mathbf{u}}(t) \right) \\
     & = & \mathbf{C} \: \dfrac{\partial \mathbf{x}(t)}{\partial A_{i \: j}} + \overline{\mathbf{D}} \: \dfrac{\partial \overline{\mathbf{u}}(t)}{\partial A_{i \: j}} \\
		& = &  \mathbf{C} \: \dfrac{\partial \mathbf{x}(t)}{\partial A_{i \: j}} + \overline{\mathbf{D}} \: \dfrac{\partial \overline{\mathbf{u}}(t)}{\partial \mathbf{y}(t)} \dfrac{\partial \mathbf{y}(t)}{\partial A_{i \: j}} .
\end{array}
\label{Eq:JWLS_A1}
\end{equation}
The first term in the right-hand side of Eq.~(\ref{Eq:JWLS_A1}) is obtained by taking the derivative of the state relation in Eqs.~(\ref{Eq:GreyBoxStateSpace}) with respect to $A_{i \: j}$, that is
\begin{equation}
\begin{array}{r c l}
    \dfrac{\partial \mathbf{\dot{x}}(t)}{\partial A_{i \: j}} & = & \dfrac{\partial}{\partial A_{i \: j}} \left ( \mathbf{A} \: \mathbf{x}(t) + \overline{\mathbf{B}} \: \overline{\mathbf{u}}(t) \right) \\
     & = & \mathbf{A} \: \dfrac{\partial \mathbf{x}(t)}{\partial A_{i \: j}} + \mathbf{I}^{\: n_{s} \times n_{s}}_{i \: j} \: \mathbf{x}(t) + \overline{\mathbf{B}} \: \dfrac{\partial \overline{\mathbf{u}}(t)}{\partial A_{i \: j}} \\
		& = &  \mathbf{A} \: \dfrac{\partial \mathbf{x}(t)}{\partial A_{i \: j}} + \mathbf{I}^{\: n_{s} \times n_{s}}_{i \: j} \: \mathbf{x}(t) + \overline{\mathbf{B}} \: \dfrac{\partial \overline{\mathbf{u}}(t)}{\partial \mathbf{y}(t)} \dfrac{\partial \mathbf{y}(t)}{\partial A_{i \: j}} ,
\end{array}
\label{Eq:JWLS_A2}
\end{equation}
where $\mathbf{I}^{\: n_{s} \times n_{s}}_{i \: j}$ is a zero matrix with a single element equal to one at entry $\left( i,j \right)$.\\

The element $J_{A_{ij}}(t)$ is therefore given by the solution of the two equations
\begin{equation}
\left\lbrace
\begin{array}{r c l}
    \dfrac{\partial \mathbf{\dot{x}}(t)}{\partial A_{i \: j}} & = & \mathbf{A} \: \dfrac{\partial \mathbf{x}(t)}{\partial A_{i \: j}} + \mathbf{I}^{\: n_{s} \times n_{s}}_{i \: j} \: \mathbf{x}(t) + \overline{\mathbf{B}} \: \dfrac{\partial \overline{\mathbf{u}}(t)}{\partial \mathbf{y}(t)} \dfrac{\partial \mathbf{y}(t)}{\partial A_{i \: j}} \\
	  \dfrac{\partial \mathbf{y}(t)}{\partial A_{i \: j}} & = & \mathbf{C} \: \dfrac{\partial \mathbf{x}(t)}{\partial A_{i \: j}} + \overline{\mathbf{D}} \: \dfrac{\partial \overline{\mathbf{u}}(t)}{\partial \mathbf{y}(t)} \dfrac{\partial \mathbf{y}(t)}{\partial A_{i \: j}} .
\end{array} \right.
\label{Eq:JAij}
\end{equation}
Introducing the notations
$$ \begin{array}{c c}
\mathbf{x^{\ast}}(t) = \dfrac{\partial \mathbf{x}(t)}{\partial A_{i \: j}} \: ; & \mathbf{y^{\ast}}(t) = \dfrac{\partial \mathbf{y}(t)}{\partial A_{i \: j}} \: ; \\
\end{array} $$
\begin{equation}
\begin{array}{c}
 \overline{\mathbf{u}}^{\ast}(t) = \left( \begin{array}{c c}
             \mathbf{x}(t)^{T} & \left( \dfrac{\partial \overline{\mathbf{u}}(t)}{\partial \mathbf{y}(t)} \dfrac{\partial \mathbf{y}(t)}{\partial A_{i \: j}} \right)^{T} \\
             \end{array} \right)^{T} \\
\end{array}
\label{Eq:JAij_Notation1}
\end{equation}
and
$$ \begin{array}{c c}
\mathbf{A^{\ast}} = \mathbf{A} \: ; & \overline{\mathbf{B}}^{\ast} = \left( \begin{array}{c c}
             \mathbf{I}^{\: n_{s} \times n_{s}}_{i \: j} & \overline{\mathbf{B}} \\
             \end{array} \right);
\end{array} $$	
\begin{equation}
\begin{array}{c c}					
						\mathbf{C^{\ast}} = \mathbf{C} \: ; & \overline{\mathbf{D}}^{\ast} = \left( \begin{array}{c c}
             \mathbf{0}^{\: l \times n_{s}} & \overline{\mathbf{D}} \\
             \end{array} \right) ,
\end{array}
\label{Eq:JAij_Notation2}
\end{equation}
Eqs.~(\ref{Eq:JAij}) can be recast in the form
\begin{equation}
\left\lbrace
\begin{array}{r c l}
    \mathbf{\dot{x}^{\ast}}(t) & = & \mathbf{A^{\ast}} \: \mathbf{x^{\ast}}(t) + \overline{\mathbf{B}}^{\ast} \: \overline{\mathbf{u}}^{\ast}(t) \\
    \mathbf{y^{\ast}}(t) & = & \mathbf{C^{\ast}} \: \mathbf{x^{\ast}}(t) + \overline{\mathbf{D}}^{\ast} \: \overline{\mathbf{u}}^{\ast}(t) .
\end{array} \right.
\label{Eq:JAij_ast}
\end{equation}
Eqs.~(\ref{Eq:JAij_ast}) reveal that the elements of the Jacobian matrix associated with the parameters in $\mathbf{A}$ are solutions of an auxiliary state-space model defined by the four matrices $\left( \mathbf{A^{\ast}},\overline{\mathbf{B}}^{\ast},\mathbf{C^{\ast}},\overline{\mathbf{D}}^{\ast} \right)$. The first term in the auxiliary extended input $\overline{\mathbf{u}}^{\ast}(t)$ in Eq.~(\ref{Eq:JAij_Notation1}) is the state vector $\mathbf{x}(t)$. It is obtained by simulating in time the original model in Eqs.~(\ref{Eq:GreyBoxStateSpace}) with the estimated parameters of the previous Levenberg-Marquardt iteration. The second term in $\overline{\mathbf{u}}^{\ast}(t)$ depends on $\partial \overline{\mathbf{u}}(t) / \partial \mathbf{y}(t)$, which is formed using the derivatives of the nonlinear basis functions $\mathbf{g}(\mathbf{y}_{nl}(t),\mathbf{\dot{y}}_{nl}(t))$ with respect to $\mathbf{y}(t)$.\\

The determination of the element $J_{\overline{B}_{ij}}(t) \in \mathbb{R}^{\: l}$ is conducted similarly to $J_{A_{ij}}(t)$. The result is given in Eqs.~(\ref{Eq:JBij}), where $J_{\overline{B}_{ij}}(t)$ is seen to be the solution of another auxiliary state-space model,

\begin{equation}
\left\lbrace
\begin{array}{r c l}
    \dfrac{\partial \mathbf{\dot{x}}(t)}{\partial \overline{B}_{i \: j}} & = & \mathbf{A} \: \dfrac{\partial \mathbf{x}(t)}{\partial \overline{B}_{i \: j}} + \mathbf{I}^{\: n_{s} \times (m + s l)}_{i \: j} \: \overline{\mathbf{u}}(t) + \overline{\mathbf{B}} \: \dfrac{\partial \overline{\mathbf{u}}(t)}{\partial \mathbf{y}(t)} \dfrac{\partial \mathbf{y}(t)}{\partial \overline{B}_{i \: j}} \\
		& & \\
     \dfrac{\partial \mathbf{y}(t)}{\partial \overline{B}_{i \: j}} & = & \mathbf{C} \: \dfrac{\partial \mathbf{x}(t)}{\partial \overline{B}_{i \: j}} + \overline{\mathbf{D}} \: \dfrac{\partial\overline{\mathbf{u}}(t)}{\partial \mathbf{y}(t)} \dfrac{\partial \mathbf{y}(t)}{\partial \overline{B}_{i \: j}} .
\end{array} \right.
\label{Eq:JBij}
\end{equation}

The computation of $J_{C_{ij}}(t) \in \mathbb{R}^{\: l}$ and $J_{\overline{D}_{ij}}(t) \in \mathbb{R}^{\: l}$ is easier because they do not involve time integration, as shown in Eq.~(\ref{Eq:JCij}) and Eq.~(\ref{Eq:JDij}), respectively,

\begin{equation}
\begin{array}{r c l}
    \dfrac{\partial \mathbf{y}(t)}{\partial C_{i \: j}} & = & \mathbf{I}^{\: l \times n_{s}}_{i \: j} \: \mathbf{x}(t) + \overline{\mathbf{D}} \: \dfrac{\partial \overline{\mathbf{u}}(t)}{\partial \mathbf{y}(t)} \dfrac{\partial \mathbf{y}(t)}{\partial C_{i \: j}} ;
\end{array}
\label{Eq:JCij}
\end{equation}

\begin{equation}
\begin{array}{r c l}
    \dfrac{\partial \mathbf{y}(t)}{\partial \overline{D}_{i \: j}} & = & \mathbf{I}^{\: l \times (m + s l)}_{i \: j} \:\overline{\mathbf{u}}(t) + \overline{\mathbf{D}} \: \dfrac{\partial \overline{\mathbf{u}}(t)}{\partial \mathbf{y}(t)} \dfrac{\partial \mathbf{y}(t)}{\partial \overline{D}_{i \: j}} .
\end{array}
\label{Eq:JDij}
\end{equation}

\end{document}